\documentclass{article} 
\usepackage[margin=3.5cm]{geometry}

\usepackage{tikz}
\usetikzlibrary{shapes,arrows,shadows}

\usepackage{enumitem}
\usepackage{subcaption}
\usepackage{amsmath}
\usepackage{amsthm}
\usepackage{dsfont}
\usepackage{comment}
\usepackage{graphicx}
\usepackage{amssymb}
\usepackage{epsfig}
\usepackage{authblk}
\usepackage[bbgreekl]{mathbbol}
\usepackage{floatrow}
\usepackage{caption}
\usepackage[font=footnotesize, labelfont=bf]{caption}
\usepackage{epstopdf}
\usepackage[all]{xy}
\usepackage[bbgreekl]{mathbbol}
\restylefloat{figure}

\usepackage[square, sort, numbers]{natbib}
\usepackage[ruled,vlined,linesnumbered]{algorithm2e}
\usepackage{scalefnt}

\begin{document}
\title{Tracking multiple moving objects in images using Markov Chain Monte Carlo}
\author{Lan Jiang \&
        Sumeetpal S. Singh
        \thanks{The authors are at the Department of Engineering, University of Cambridge, United Kingdom.}
        \thanks{This work was supported by the Engineering and Physical Sciences Research Council [grant numbers EP/G037590/1, EP/K020153/1.]}
        }


\maketitle


\begin{abstract}
A new Bayesian state and parameter learning algorithm for multiple target tracking (MTT) models with image observations is proposed.  Specifically, a Markov chain Monte Carlo algorithm is designed to sample from the posterior distribution of the unknown number of targets, their birth and death times, states and model parameters, which constitutes the complete solution to the tracking problem. The conventional approach is to pre-process the images to extract point observations and then perform tracking. 
We model the image generation process directly to avoid potential loss of information when extracting point observations. Numerical examples show that our algorithm has improved tracking performance over commonly used techniques, for both synthetic examples and real florescent microscopy data, especially in the  case of dim targets with overlapping illuminated regions.  
\end{abstract}

\section{Introduction}
The multiple target tracking (MTT) problem is to infer the states or tracks of multiple moving objects from noisy measurements. The problem is difficult since the number of targets is unknown and changes over time as it is a birth-death process. Other compounding factors include the non-linearity of both the target's motion and observation models. 
In many applications such as radar/sonar tracking \citep{Bar-Shalom_and_Fortmann_1988} and Fluorescence Microscopy \cite{weimann2013quantitative}, the measurements (or observations) are images. (For example, a pixel's illumination intensity is a measure of nearby targets energy and background noise.) These images are usually pre-processed prior to actual tracking to extract point measurements where each point is a spatial coordinate, which are then assumed to be either noisy measurements of the target state or spuriously generated. The latter is an artefact of the method that extracts point measurements. Converting images to point measurements is advantageous because it yields a simpler observation model and also simplifies the design of tracking algorithms \citep{Bar-Shalom_and_Fortmann_1988, mahler2007statistical}, 
e.g.\ \citep{weimann2013quantitative} connects the point measurements using a nearest neighbour method to form target trajectories.
However,  the pre-processing step can introduce information loss in the low signal-to-noise (SNR) regime, which can be near complete as the targets become more closely spaced and background noise intensifies. In low SNR it can be difficult to isolate bright regions in the image, whose centres would be the candidate point measurements, and then attribute them to distinct targets.  Thus, MTT algorithms that  work with the images directly can be preferable and there is a sizeable literature on it. A selection of works is \citep{streit2002multitarget, RGM05, boers2004multitarget, punithakumar2005sequential, samuel2007comparison, vo2010joint, PaK15} and they differ in how tracking is achieved (Bayesian, maximum likelihood or otherwise) and the specific assumptions imposed on the image model.

Given images recorded over a length of time, say from time $1$ to $n$, our aim is to jointly infer the target tracks and the MTT model parameters.  We adopt a Bayesian approach and one of our main contributions is the design of a new Markov chain Monte Carlo (MCMC) algorithm for an image measurement model that can jointly track and calibrate. The MCMC algorithm is a trans-dimensional sampler that combines Particle Markov Chain Monte Carlo (PMCMC) steps \citep{andrieu2010particle} to sample from the \emph{exact} MTT posterior distribution for entire target tracks and model parameters. This is in contrast to numerous techniques that use specific variational approximations, e.g.\ spatial Poisson, of the MTT posterior to simplify inference \cite{mahler2007statistical}. Entire tracks, as opposed to point estimates of target locations at each time \cite{vo2010joint}, are needed as these are then used to infer the aggregate diffusion characteristics of the molecules \cite{weimann2013quantitative}. 
Model calibration is needed because in real fluorescence microscopy data, the molecules to be tracked bleach over time and the noise characteristics of the acquired images also drift. These changes can be captured by time varying image model parameters, which are indeed unknown to the analyst, as are other parameters such as those describing the molecule motion model. To the best of our knowledge, our trans-dimensional MCMC tracker for image observations addresses this practical tracking problem in greater generality without major limiting assumptions like well separated molecules with non-overlapping illumination regions \cite{vo2010joint} or aforementioned principled simplifications of the MTT posterior \cite{PaK15}.  In numerical examples we demonstrate the superior performance of our method over \cite{vo2010joint} for closely spaced targets. We also apply it to real florescence microscopy data  and show it outperforms a method currently used by biologists \cite{weimann2013quantitative} which pre-processes the images to extract point observations. Such comparisons, which are absent in the literature, highlight the gain in performance by targeting the exact MTT posterior and avoiding simplifications like disallowing overlaps. 

We do not advocate that our MCMC technique should replace techniques that extract point observations, those that use variational approximations to simplify the posterior \cite{mahler2007statistical, vo2010joint, PaK15}, or those that do not extract point observations but are optimised for non-overlapping targets \cite{vo2010joint}. Our MCMC technique should be viewed as a compliment to these other techniques. It could be applied in an online tracking scenario by processing a window of data at a time or as a post-processing tool to refine the trajectories identified by any other online algorithm \cite{jiang_fusion2014}. This is similar to the role MCMC plays in the related field of Particle Filtering, which is an online estimation method, where MCMC is used to refine the online estimates \cite{kantas2015}.

There is a growing literature on using MCMC for tracking as it is recognised that sampling the true MTT posterior, although challenging, is feasible in offline applications and can serve as a track refinement tool in the online setting \cite{Oh_et_al_2009, vu2014particle}. There exists several MCMC based MTT algorithms for point observation models. \cite{Oh_et_al_2009,  KoS2015} assume the underlying state-space and observation model is linear and  Gaussian, \cite{vu2014particle, Lan2014} consider the non-linear and non-Gaussian setting while \citep{Duckworth:EECS-2012-68, Lan2014,  KoS2015} simultaneously estimate the model parameters. (Although some of the above works incorporate parameter estimation, it is a topic in MTT that has only recently gained attention, see \cite{SWG11, yildirim2014calibrating}.)
Tracking using images are also known as track-before-detect (TBD) techniques. \cite{punithakumar2005sequential, vo2010joint,PaK15} use specific but different Poisson approximations for the MTT posterior (assuming known model parameters) which is then approximated using a Particle Filter. 

The remainder of the paper is organised as follows. Section \ref{sec: Multiple target tracking model} describes the MTT model and presents the framework for joint state and parameter learning algorithm. In Section \ref{sec: MCMC for z}, we present details of our novel MCMC kernel for detecting and maintaining tracks, which constitutes the core part of our tracking algorithm. (More detailed derivations are given in the Appendix.) Section \ref{sec: Numerical examples} presents numerical results for both synthetic and real florescent microscopy data.

\section{Multiple target tracking model} \label{sec: Multiple target tracking model}
\subsection{The single target model}
We commence with a description of the image based tracking problem assuming a single target and then enlarge the model for the multi-target case.
Let the Markov process $\{ X_{t} \}_{t \geq 1}$ represent the state values of a single evolving target. 
In this work it is assumed that $X_{t}=(X_t(1), \ldots, X_t(5)) \in \mathbb{R}^{5}$ where $X_t(i)$ denotes its $i$th component. $X_t(1)$ is the target's \emph{illumination intensity} or amplitude (to be discussed in detail next), $(X_t(2),X_t(3))$ is the spatial coordinate of the target and $(X_t(4),X_t(5))$ are the corresponding spatial velocities. Frequent reference will be made to the intensity (amplitude), spatial coordinate and spatial velocity components of a target state $X_t$. As such we will denote these components by $A_t=X_t(1)$, $S_t=(S_t(1),S_t(2))=(X_t(2),X_t(3))$ and $V_t=(V_t(1),V_t(2))=(X_t(4),X_t(5))$.  $X_{t}$ is a time-homogeneous Markov process,
\begin{equation} \label{eq:state}
X_{1} \sim \mu_\psi(\cdot), \quad X_{t} |X_{1:t-1} = x_{1:t-1} \sim f_{\psi} ( \cdot | x_{t-1})
\end{equation}
where  $\mu_{\psi}$ and $f_{\psi}$ are, respectively,  the initial and state transition probability density function (pdf), both parametrised by the common real valued vector $\psi \in \Psi \subset \mathbb{R}^{d_{\psi}}$. (As a rule, a random variable (r.v.) is denoted by a capital letter and its realisation by small case.) For example, for linear and Gaussian state dynamics,  $\mu_{\psi}(x)=\mathcal{N}(x; \mu_b, \Sigma_b)$, $f_{\psi}(x'|x)=\mathcal{N}(x'; Fx, W)$, where $\mathcal{N}(\cdot; m, \Sigma)$ denotes the Gaussian pdf with mean $m$ and covariance matrix $\Sigma$. Thus $\psi=(\mu_b,\Sigma_b,F,W)$. 

For the measurements, a two dimensional image measurement model is assumed with $m$ pixels in total. Let \[Y_t=(Y_{t,1},\ldots, Y_{t,m}),\] 
denote the observed image at time $t$ where $Y_{t,i}$ is the value (illumination intensity) of pixel $i$.
$Y_{t,i}$ is  defined as
\begin{equation} Y_{t,i}=h_i(X_{t})+E_{t,i}, \label{Eq:PointObs_0}\end{equation}
where $E_{t,i}$ is the observation noise of pixel $i$ at time $t$ and $h_i(X_{t})$ is the illumination of pixel $i$ by a single target with state $X_t$. 
As in \cite{vo2010joint}, for  $x=(a,s,v) \in \mathbb{R}\times \mathbb{R}^2 \times \mathbb{R}^2$ where $a$ is the intensity, $s=(s(1),s(2))$ the spatial coordinate and $v=(v(1),v(2))$ the spatial velocity, $h_i(x)$  is the point spread function 
\begin{align}
h_i(x) &= \mathbb{I}[i\in L(s)] \; \frac{a \Delta_1\Delta_2}{2\pi\sigma_h^2} \nonumber \\
& \quad \times  \exp\{-\frac{(\Delta_1 r -s(1))^2+(\Delta_2 c-s(2))^2}{2\sigma_h^2}\}  \nonumber \\
&=:  \;  a \; \bar{h}_i(s)
\label{Eq:PointObs}
\end{align} 
where $(r,c)$ denotes the row and column number of pixel $i$, $\Delta_1$  and $\Delta_2$ are constants that map pixel indices to spatial coordinates and $\sigma_h$  is the blurring parameter. It is assumed that the spatial coordinate of the pixel with index corresponding to row and column number $(0,0)$ is the origin of $\mathbb{R}^2$. As in  \cite{vo2010joint},  we also assume for each state value $x=(a,s,v)$ there is a square truncation region $L(s)$  where $h_i(x)=0$ if $ i\notin L(s)$. Specifically, $L(s)$ is the set of $l \times l$ pixels, $l$ an odd integer, whose centre pixel has spatial coordinate closest to $s$. Henceforth we assume $\Delta_1=\Delta_2=\Delta$.

For later use, the function $\bar{h}_i(s)$ in \eqref{Eq:PointObs} has been implicitly defined. In addition, extend the domain of the truncation region $L$ and point spread function
$\bar{h}_i$ to included pixel indices $j\in\{1,\ldots,m\}$. That is, let $(r',c')$ be the row and column number of pixel $j$  and define
\begin{equation}
L(j)=L(s), \quad \bar{h}_i(j) = \bar{h}_i(s) \quad \textrm{where } s=(r' \Delta, c' \Delta).
\label{eq:psfandL}
\end{equation}
Equivalently $L(j)$ is the square of $l \times l$ pixels centered at pixel $j$.

The pixel noise is assumed to be Gaussian with mean value $b_t$, representing the background intensity, and variance $\sigma_{r,t}^2$, both time varying but common across pixels, i.e.\footnote{$\sigma_{r,t}$ is the symbol for the observation noise and subscript $r$ is not to be confused with row number mentioned before.}
\[E_{t,i}\overset{\text{i.i.d.}}{\sim} \mathcal{N}(\cdot|b_t, \sigma_{r,t}^2), \quad i=1,\ldots,m, \; t=1,\ldots,n. \] 
Thus, the conditional pdf of the observed image at time $t$ due to a single target with state $X_t$  is 
\begin{equation*}
g_t(y_{t}|x_t)
= \prod_{i=1}^m\mathcal{N}(y_{t,i}; h_i(x_t)+b_t, \sigma_{r,t}^2).
\end{equation*}
where subscript $t$ of $g_t$ indicates the observation model is time-inhomogeneous. 
Given $n$ images, all the model parameters $(\psi,b_1, \sigma_{r,1},\ldots, b_n, \sigma_{r,n})$ described in this section will be estimated.

\subsection{The model for multiple targets}
In this section we partially adopt the formulation in \cite{Lan2014} for the MTT model. (Note though that the observation model in \cite{Lan2014} is for point-observations and not for images as in our case.) In an MTT model,  the MTT state at time $t$ is the concatenation of all individual target states at $t$:
\[
\mathbf{X}_{t} = \left( X_{t, 1}, X_{t, 2}, \ldots, X_{t, K^{x}_{t}} \right)\]
where each sub-vector $X_{t,i}$ is the state (as in \eqref{eq:state}) of an individual target. The number of targets $K^{x}_{t}$ under surveillance changes over time due to the death of existing targets and the birth of new targets.  Independently of the other targets, a target survives to the next time with survival probability $p_{s}$ and its state evolves according to the transition density $f_{\psi}$, otherwise it `dies'. In addition to the surviving targets, new targets are `born' from a Poisson process with intensity $\lambda_{b}$ and each of their states is initialised by sampling from the initial density $\mu_{\psi}$. The  states of the new born targets and surviving targets from time $t$ make up $\mathbf{X}_{t+1}$. We assume that at time $t=1$ there are only new born targets, i.e.\ no surviving targets from the past.

\begin{figure}
\centering
\includegraphics[width=.5\linewidth]{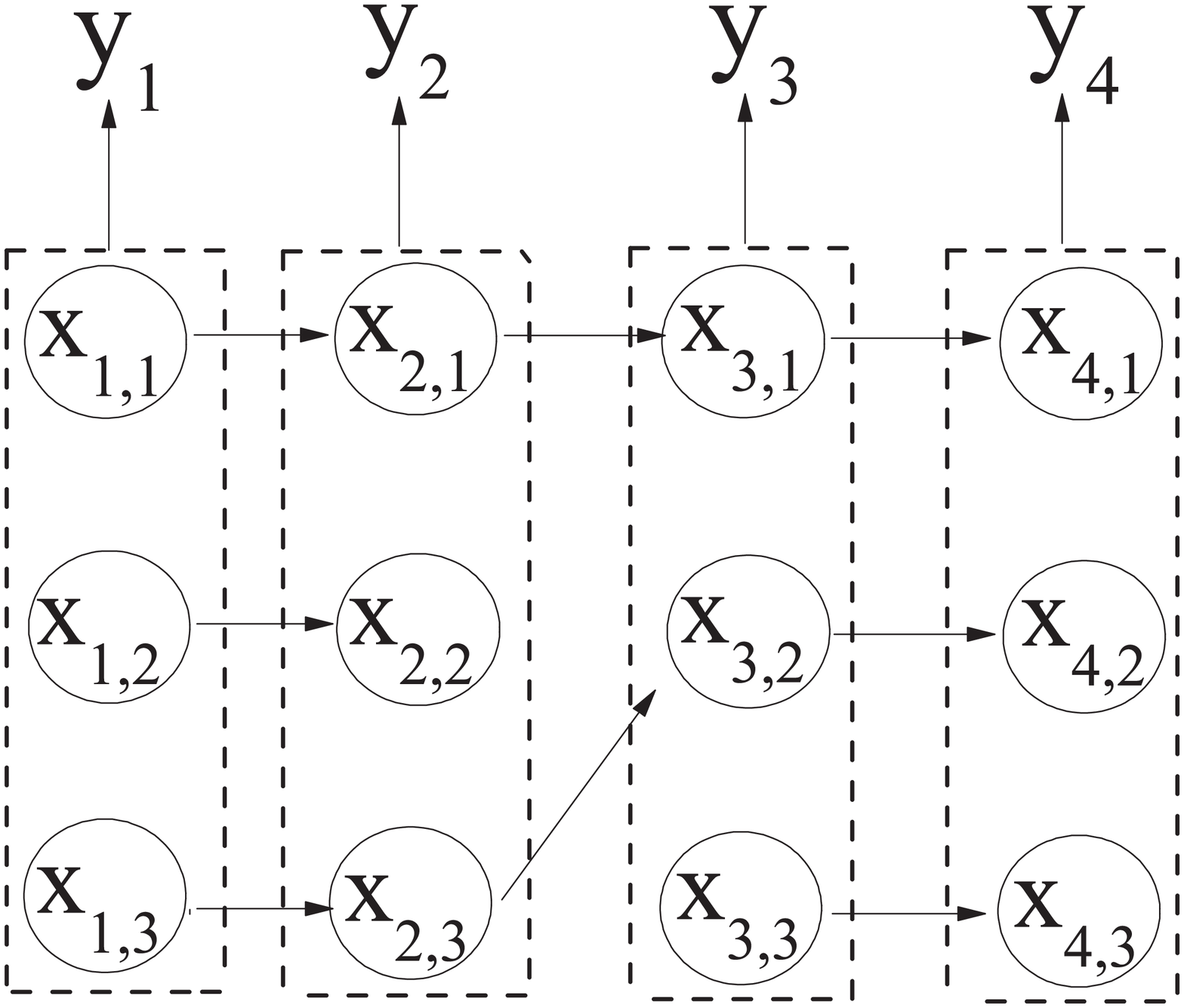}
\caption{A realisation from  the  MTT model: states of a target are connected with arrows and all targets at time $t$ contribute to image $y_t$.\\
	\textbf{MTT random variables:} \\
	Time $\mathit{t = 1}$: No prior targets ($C_{1} = (), K^{s}_{1} = 0, I_{1} = ()$), three targets are born ($K^{x}_{1} = K^{b}_{1} = 3$) with states $X_{1, 1}, X_{1, 2}, X_{1, 3}$;  \\
	Time $\mathit{t = 2}$: All targets $X_{1, 1}, X_{1, 2}, X_{1, 3}$ survive to $X_{2, 1}, X_{2, 2}, X_{2, 3}$. Thus $C_{2} = (1, 1, 1)$, $K^{s}_{2} = 3$, $I_{2} = (1, 2, 3)$. No new born targets, $K^{b}_{2} = 0$, $K^{x}_{2} = K^{s}_{2} + K^{b}_{2} = 3$.\\
	Time $\mathit{t = 3}$: Targets $X_{2, 1}$ and $X_{2, 3}$ survive to $X_{3, 1}$ and $X_{3, 2}$ respectively while $X_{2, 2}$ dies, thus $C_{3} = (1, 0, 1)$, $K^{s}_{3} = 2 $, $I_{3} = (1, 3)$. One new born target, $K^{b}_{3} = 1$, denoted $X_{3, 3}$.  $K^{x}_{3} = K^{s}_{3} + K^{b}_{3} = 3$. \\
	Time $\mathit{t = 4}$: All targets survive, no new born, same as time $t=2$.  \\
	\textbf{MTT variables of the equivalent description of Sec.\ \ref{sec: Two equivalent notations for MTT}:} 
	\newline $t_b^{1}=1$, ${\mathbf{X}}^{1}=(X_{1,1}, X_{2,1}, X_{3,1}, X_{4,1})$;
	$ t_b^{2}=1$, ${\mathbf{X}}^{2}=(X_{1,2}, X_{2,2})$; $ t_b^{3}=1$,  ${\mathbf{X}}^{3}=(X_{1,3}, X_{2,3}, X_{3,2}, X_{4,2})$;     
	$t_b^{4}=3$, ${\mathbf{X}}^{4}=(X_{3,3}, X_{4,3})$.
	}
\label{fig: MTT}
\end{figure}

To describe the evolution  of $\mathbf{X}_t$ due to  survivals and  births, a series of random variables are now defined here.
Let $K_t^x$ and  $K_t^b$ denote the number of targets and new births at time $t$ respectively. We start with $K_1^x=K_1^b$. For $t>1$ and $ i = 1,\ldots,K^{x}_{t-1}$, let
\begin{equation}
C_{t}(i) = \begin{cases}
      1 & \text{$i$th target at time $t-1$ survives to time $t$} \\
      0 & \text{$i$th target at time $t-1$ does not survive to $t$}
\end{cases}. \nonumber
\end{equation}
$C_{t}$ is the $K^{x}_{t-1} \times 1$ binary  vector where $1$'s indicate survivals and $0$'s indicate deaths of targets from time $t-1$.
Let  $K_t^s$ denote  the number of surviving targets at time $t$, thus  \[K^{s}_{t} = \sum_{i = 1}^{K^{x}_{t-1}} C_{t}(i).\]
The $K^{s}_{t}$ surviving targets from time $t-1$ evolve to become the \emph{first} $K^{s}_{t}$ targets in $\mathbf{X}_{t}$. Specifically, define the $K^{s}_{t} \times 1$ ancestor vector $I_{t}$, $t>1$, as
\begin{equation}
I_{t}(i) = \min \bigl\{ k: \sum_{j = 1}^{k} C_{t}(j) = i \bigr\}, \quad i = 1, \ldots, K^{s}_{t}, \nonumber
\end{equation}
so that $X_{t-1, I_{t}(i)}$ evolves to $X_{t, i}$ for $i = 1,\ldots,K^{s}_{t}$. In addition to the surviving targets $(X_{t, 1}, \ldots, X_{t, K^{s}_{t}})$, we have $K^{b}_{t}$ newly born targets denoted by $X_{t, K^{s}_{t}+1}, \ldots, X_{t, K^{x}_{t}}$. The state  $\mathbf{X}_{t}$ is formed of   the new born targets together with the surviving targets, and thus $K_t^x=K_t^b+K_t^s$. An ordering rule is adopted for the new born targets to avoid labelling ambiguity. Specifically, the new born targets at each time $t$ are labelled in ascending order of their first component value.
Let $Z_1=K_1^b$ and 
\[Z_{t} = \left( C_{t}, K_t^b \right ), \quad t>1,\] which is the discrete component of the MTT state at time $t$. Figure \ref{fig: MTT} illustrates all the MTT random variables.

\subsection{The law of MTT model}\label{subsec:MTT_model}

The image observation $Y_t=(Y_{t,1},\ldots, Y_{t,m})$ generated by multiple targets at time $t$ is the  superposition of  the contributions of all targets at time $t$, the background intensity and noise, i.e.
\begin{equation}
Y_{t,i}=h_i(\mathbf{X}_t)+E_{t,i},\quad  h_i(\mathbf{X}_t)=\sum_{k=1}^{K_t^x}h_i(X_{t,k}),
\label{eq:ObsModel}
\end{equation}
where  $h_i(X_{t,k})$ is the contribution of the $k$-th target at time $t$ to the illumination of pixel $i$ (see \eqref{Eq:PointObs}). The MTT observation  model is 
\begin{equation}
g_{t}(y_{t}|\mathbf{x}_t) = \prod_{i=1}^m\mathcal{N}(y_{t,i}; h_i(\mathbf{x}_t)+b_t, \sigma_{r,t}^2). \label{eq:mttobs}
\end{equation}
Given the vector of the MTT model parameters
\begin{equation}
\theta = (\psi,  p_{s},\lambda_{b}, b_1, \sigma_{r,1}^2,\ldots,b_n, \sigma_{r,n}^2)\label{eq:theta}
\end{equation}
the law of the MTT model can be expressed with the joint density of  $(Z_{1:n}, \mathbf{X}_{1:n}, {Y}_{1:n})$ which is
\[
p_{\theta}(z_{1:n}, \mathbf{x}_{1:n}, y_{1:n}) =   p_{\theta}(y_{1:n} | \mathbf{x}_{1:n}, z_{1:n})  p_{\theta}(\mathbf{x}_{1:n} | z_{1:n}) p_{\theta}(z_{1:n})
\]
where  $a_{i:j}$, $i\leq j$, denotes the sequence $a_i,a_{i+1}\ldots a_j$.
\begin{align}
& p_{\theta}(y_{1:n}|\mathbf{x}_{1:n},z_{1:n})=\prod_{t=1}^n g_{\theta} (y_{t}|\mathbf{x}_t),\label{eq: density of y given x,z}\\
& p_{\theta}(z_{1:n}) = \mathcal{P}(k^{b}_{1} ; \lambda_{b}) \prod_{t = 2}^{n}  p_{s}^{k^{s}_{t}} (1 - p_{s})^{k^{x}_{t-1} - k^{s}_{t}}  \mathcal{P}(k^{b}_{t} ; \lambda_{b}), \label{eq: density of z}
\end{align}
\begin{align}
&p_{\theta}(\mathbf{x}_{1:n} | z_{1:n} ) = \prod_{t = 1}^{n}\, \biggl[ \prod_{ j=1 }^{k_t^s} f_{\psi}(x_{t, j} | x_{t-1, i_{t}(j)})\nonumber \\
&\hspace{2.5cm} k_t^b! \mathbb{I}_{\mathcal{O}}(x_{t,k_t^s+1:k_t^x})\prod_{j=k_t^s+1}^{k_t^x}\mu_{\psi}(x_{t, j}) \biggr]\label{eq: density of x given z}
\end{align}
where  $\mathcal{P}(k; \lambda)$ denotes the probability mass function of the Poisson distribution with mean $\lambda$ and  $g_{\theta} (y_{t}|\mathbf{x}_t)$ in \eqref{eq: density of y given x,z} is only dependent on components $(b_t, \sigma_{r,t})$ of $\theta$ and is precisely $g_t$ of \eqref{eq:mttobs}. In \eqref{eq: density of x given z}, $\mathbb{I}_{\mathcal{O}}$ is the indicator function of the particular ordering rule $\mathcal{O}$ for the new born targets,
\[
\mathbb{I}_{\mathcal{O}}(x_{t,k_{t}^{s}+1:k_{t}^{x}}) =  \begin{cases}
1 & \text{if $x_{t,k_t^s+1}(1)<\cdots<x_{t,k_t^x}(1)$}, \\
0 & \text{else}.
\end{cases}
\] 
Note that ordering the latent variables is also done in other statistical problems where labelling ambiguity arises through the likelihood function, e.g.\ Bayesian inference of mixture distributions \citep{jasra2005markov}. For MTT, the ordering will 
be very useful in Section \ref{sec: Two equivalent notations for MTT} where, thanks to rule $\mathcal{O}$, we are able to deterministically assign a unique label to each target track.
Finally, the marginal likelihood of the data $y_{1:n}$ is given by
\[
p_{\theta}(y_{1:n})\! =\! \sum_{z_{1:n}} p_{\theta}(z_{1:n})\hspace{-0.1cm} \int\hspace{-0.1cm} p_{\theta}(y_{1:n} | \mathbf{x}_{1:n}, z_{1:n}) p_{\theta}(\mathbf{x}_{1:n} | z_{1:n}) \textrm{d} \mathbf{x}_{1:n}.
\]

\subsection{An equivalent representation of $(Z_{1:n}, \mathbf{X}_{1:n})$} \label{sec: Two equivalent notations for MTT}
This section introduces an equivalent parameterization for the MTT problem. Essentially, we define a new set of random variables which are an alternative to those defined in Section \ref{subsec:MTT_model} without any loss of information. The idea here is to introduce notation that explicitly isolates the state trajectories of individual targets, which will be very useful to describe the MCMC proposal distributions in Section \ref{sec: MCMC for z}.

Let $K = \sum_{t = 1}^{n} k_{t}^{b}$ denote the total number of targets that have appeared from time $1$ to $n$. 
Each target appearing in this time span can be assigned a distinct label or index  $k\in\{1,\ldots, K\}$ with the convention that targets born earlier are given a smaller label than those born at a later time and targets born at the same time are sorted by the ordering rule $\mathcal{O}$.

Consider a target assigned labelled $k\in\{1,\ldots, K\}$, let it's birth time be $t_{b}^{k}$, death time be $t_{d}^{k}$
and its life span be $l^{k} = t_{d}^{k} - t_{b}^{k}$. (Note $t_{d}^{k}-1$ is the final time of its existence.) The entire continuous state trajectory of this target can be extracted from the MTT state sequence
$(\mathbf{X}_{t_{b}^{k}},\ldots,\mathbf{X}_{t_{d}^{k}-1})$ and denote it by
\[
{\mathbf{X}}^{k}=({X}_{0}^{k},\ldots, {X}_{l_k-1}^{k})\]
where  ${X}_{i-1}^{k}$ is the $i$-th state of target $k$. Note that  ${\mathbf{X}}^{k}$ is a Markov process with initial and state transition densities $\mu_{\psi}$ and $f_{\psi}$ respectively. It is straightforward to extract $\{ (k,t_{b}^{k},{\mathbf{X}}^{k}) \}_{k=1}^K$ from $(Z_{1:n}, \mathbf{X}_{1:n})$ as illustrated in Figure \ref{fig: MTT}. The main point is that we can use one of the two equivalent descriptions for  latent variables of the MTT model, i.e.
\begin{equation}
(Z_{1:n}, \mathbf{X}_{1:n}) \Leftrightarrow \{ (k, t_{b}^{k},{\mathbf{X}}^{k}) \}_{k=1}^K.\label{eq:TwoNotation}
\end{equation}
On the other hand, $(Z_{1:n}, \mathbf{X}_{1:n})$ can be obtained from $\{ (k, t_{b}^{k}, {\mathbf{X}}^{k})\}_{k=1}^K$ since the underlying transformation is a bijection. (Again see Figure \ref{fig: MTT} for an example.)

\subsection{Bayesian tracking and parameter estimation for MTT} \label{sec: Bayesian smoothing and parameter estimation for MTT}
The inference task is to estimate the discrete variables $Z_{1:n}$,  target states $\mathbf{X}_{1:n}$ and the MTT parameter $\theta$ given the observations $y_{1:n}$.  Regarding $\theta$ as a random variable taking values in $\Theta$ with a prior density $\eta(\theta)$, the goal is to obtain Monte Carlo samples from
\begin{equation}
 p(z_{1:n}, \mathbf{x}_{1:n}, \theta | y_{1:n}) \propto \eta(\theta) p_{\theta}(z_{1:n}, \mathbf{x}_{1:n},y_{1:n}). \label{eq:MTTposterior}
\end{equation}
We achieve this by iteratively performing  the MCMC sweeps given in Algorithm \ref{alg: MCMC for static parameter estimation}. A single call of Algorithm \ref{alg: MCMC for static parameter estimation} will transform a current sample $(\theta, Z_{1:n}, \mathbf{X}_{1:n})$ from the posterior to a new sample $(\theta', Z'_{1:n}, \mathbf{X}'_{1:n})$. The entire sequence of samples yielded by the repeated calls to Algorithm \ref{alg: MCMC for static parameter estimation} will constitute the desired set of Monte Carlo samples from \eqref{eq:MTTposterior}. We need though to discard an initial sequence of this set so that the remaining samples retained are correctly distributed. 
\begin{algorithm}[h]
	\DontPrintSemicolon
	\KwIn{Current sample $(\theta, z_{1:n}, \mathbf{x}_{1:n})$, data $y_{1:n}$, number of inner loops $n_{1}$, $n_{2}$, $n_{3}$.}
	\KwOut { Updated sample $(\theta', z'_{1:n}, \mathbf{x}'_{1:n}$).}
	\For{$j = 1:n_{1}$}
	{Update $(z_{1:n}, \mathbf{x}_{1:n})$ by invoking Algorithm \ref{alg: MCMC across target configurations}.} 
	Isolate target trajectories (see \eqref{eq:TwoNotation}) $\{ (t_{b}^{k},{\mathbf{x}}^{k}) \}_{k=1}^K.$\\
	\For{$j = 1:n_{2}$}{
	\For {$k=1:K$}
	{Update ${\mathbf{x}}^{k}$ using particle Gibbs conditioned on other trajectories ($\neq k$) and $\theta$.
	} 
	}
	Call the updated sample $(z'_{1:n}, \mathbf{x}'_{1:n}$). \\
	\For{$j = 1:n_{3}$}
	{Conditioned on $(z'_{1:n},x'_{1:n})$, update $\theta$ to $\theta'$ using a MH move or a Gibbs move.}
	\caption{MCMC for state and parameter learning}\label{alg: MCMC for static parameter estimation}
\end{algorithm}
Section~\ref{sec: MCMC for z} is dedicated to the exposition of the first loop of Algorithm~\ref{alg: MCMC across target configurations}. The remaining two loops are more easily described. The principal aim of the second loop is to resample the continuous state trajectory of each target using the particle Gibbs sampler \cite{andrieu2010particle} (but using the implementation in \cite{whiteleydiscussion}) which we find enhances our MCMC algorithm's efficiency significantly. This is done by first explicitly isolating the state trajectories of individual targets as in Section \ref{sec: Two equivalent notations for MTT} and then updating the targets' trajectories in turn using the Particle Gibbs sampler.
When conjugate priors are available for the components of $\theta$, it is possible to sample $p(\theta | z_{1:n},  \mathbf{x}_{1:n}, y_{1:n})\propto p(\theta)p_{\theta}(z_{1:n}, \mathbf{x}_{1:n}, y_{1:n})$ exactly in the final loop, in which case set $n_{3} = 1$. Otherwise, one can run a Metropolis-Hastings (MH) algorithm to sample from this pdf. When the MTT parameters are known, the third loop can be omitted and we refer to the resulting algorithm as the \textit{MCMC tracker}.

\section{MCMC moves} \label{sec: MCMC for z}
In this section, we present the MCMC moves to explore $(Z_{1:n},\mathbf{X}_{1:n})$ for the first loop in Algorithm~\ref{alg: MCMC for static parameter estimation}.
Notice that the dimension of $\mathbf{X}_{1:n}$, which is proportional to $\sum_{t = 1}^{n} K^{x}_{t}$, changes with $Z_{1:n}$. Therefore, the posterior distribution $p_{\theta}(z_{1:n}, \mathbf{x}_{1:n}|y_{1:n})$ is said to be \emph{trans-dimensional}.

\subsection{A brief on trans-dimensional MCMC} \label{sec:rjmcmcreview}
A general method for sampling from a trans-dimensional distribution is the reversible jump MCMC (RJMCMC) algorithm of \cite{green1995reversible}. We briefly describe RJMCMC  for the (general) target distribution  $\pi(m, x_{m})$ where $m$ is a discrete variable (e.g. $m\in \{1,2\ldots\}$) known as the model index and $x_m \in \mathbb{R}^{d_{m}}$. Note though that in general  $m' \neq m$ does not imply $d_{m'} \neq d_{m}$. We now define the procedure for generating samples from a different dimension and either accepting or rejecting them.

For each $(m, x_m)$, let $Q(m'| m, x_m)$ be a probability mass function satisfying $\sum_{m'} Q(m'|m,x)=1$ and $Q(m'| m, x_m)=0$ if $d_{m'}=d_m$. Furthermore, for each $m'$ such that $d_{m'}>d_m$, let $Q(u| m, x_m,m')$ be a pdf on $\mathbb{R}^{d_{m'}-d_m}$. $Q$ will be the proposal distribution and $Q(m'| m, x_m)=0$ if $d_{m'}=d_m$ implies $Q$ only proposes moves across dimension. (Sampling from $\pi$ can be achieved by combining this proposal with others that do not move across dimensions and then cycling between them either randomly or deterministically. The derivation of the accept-reject probability is routine for intra-dimensional moves.) 

Let $(m, x_{m}) \sim \pi $. First sample $m'$ from $Q(\cdot| m, x_m)$  and if $d_{m'}>d_m$, then sample $u$ from $Q(\cdot| m, x_m,m')$, which are the extra (so called dimension matching) continuous r.v.'s needed to generate the candidate sample $x_{m'} \in \mathbb{R}^{d_{m'}}$. (Indeed it is equivalent to jointly sample $(m',u)$, as opposed to doing it one after the other. The order is application dependent.)
Assume  $d_{m'}>d_m$. (The reverse case is considered below.) The candidate sample is obtained by applying a \textit{bijection} (to be chosen by the practitioner just as $Q$ was) to yield $x_{m'}=\beta_{m,m'}(x_m, u)$. 
The acceptance probability for the proposed sample $(m', x_{m'})$ is $\alpha(m',x_{m'}; m, x_m)=\min\{1, r(m',x_{m'}; m, x_m)\}$  where $r(m',x_{m'}; m, x_m)$ is
\begin{equation}
	\frac{\pi(m', x_{m'})}{\pi(m, x_{m})}
	\frac{Q(m | m', x_{m'})}{ Q(m' , u | m, x_m) }
	\!\!\left\vert  \nabla \beta_{m,m'}(x_{m}, u)    \right \vert \label{eq:TransMC}
\end{equation}
where the right most term is the Jacobian of $\beta_{m,m'}$. (If $x_{m'}=\beta_{m,m'}(x_m, u)$ is a mapping that permutes the components of the vector $(x_m, u)$ then the Jacobian is 1.) If however  $d_{m'}<d_m$, let $(x_m',u)=\beta_{m',m}^{-1}(x_m)$ and the candidate sample for the move to the lower dimension model is $x_m'$. The proposal $(m',x_{m'})$ is accepted with probability $\min\{1,r\}$ where $r(m',x_{m'}; m, x_m)$ is
\begin{equation}
\frac{\pi(m', x_{m'})}{\pi(m, x_{m})}
	\frac{Q(m,u | m', x_{m'})}{ Q(m' | m, x_m) }
	\!\!\left\vert  \nabla \beta_{m',m}(x_{m'}, u)    \right \vert^{-1} \label{eq:TransMC_reverse}
\end{equation}

In the MTT model, each target configuration $z_{1:n}$  corresponds to a model index $m$, $\mathbf{x}_{1:n}$ corresponds to the continuous variable $x_m$, and $p_{\theta}(z_{1:n}, \mathbf{x}_{1:n}|y_{1:n})$ corresponds to $\pi(m, x_m)$. The bijections $\beta_{m,m'}$ will do nothing more than permute the input variables to preserve the MTT ordering rule $\mathcal{O}$ and thus all Jacobians are 1.

\subsection{MCMC to explore $(Z_{1:n},\mathbf{X}_{1:n})$ in loop 1 of Algorithm \ref{alg: MCMC for static parameter estimation}}\label{sec: across DA}

Algorithm \ref{alg: MCMC across target configurations}  proposes a change to $(Z_{1:n},\mathbf{X}_{1:n})$ by selecting one of the following proposals at random:
\begin{enumerate}
\item \emph{birth/death} proposal to create or delete a target;
\item multi-step \emph{extension/reduction}  proposal to extend/reduce an existing track by multiple time units;
\item one-step \emph{extension/reduction} proposal to extend/reduce an existing track by one time unit;
\item \emph{state} proposal to exchange states between targets.
\end{enumerate}
All but the state proposal changes the dimension of $\mathbf{X}_{1:n}$ and hence the corresponding RJMCMC acceptance probability in \eqref{eq:TransMC} needs to be derived. However the bijections are such that the Jacobian in \eqref{eq:TransMC} is always $1$. Each proposal type is now described in detail in the subsections below 

\begin{algorithm}[h]
\DontPrintSemicolon
\KwIn{Sample $(z_{1:n}, \mathbf{x}_{1:n})$, data $y_{1:n}$, parameter $\theta$} 
\KwOut {Updated sample $(z'_{1:n},\mathbf{x}'_{1:n})$}
Randomly select a proposal type from $\{\textrm{birth/death}, \textrm{multi-step extension/reduction}, \textrm{state}, $ 
$\textrm{one-step extension/reduction}\}$.\\
Propose $(z'_{1:n},\mathbf{x}'_{1:n})$ by executing the chosen proposal. \\
Calculate acceptance prob. 
$\alpha(z'_{1:n}, \mathbf{x}'_{1:n}; z_{1:n}, \mathbf{x}_{1:n})$ (see \eqref{eq: AcptRatio_BM_Img}/\eqref{eq: AcptRatio_DM}, 
\eqref{eq:mstep_ext_red}/\eqref{eq:mstep_ext_red2}, \eqref{eq: AcptRatio_SM_Img}).
 Output $(z'_{1:n},\mathbf{x}'_{1:n})$ with prob.\ $\alpha$, otherwise $(z_{1:n},\mathbf{x}_{1:n})$.
\caption{MCMC moves to explore $(Z_{1:n},\mathbf{X}_{1:n})$} \label{alg: MCMC across target configurations}
\end{algorithm}

\subsection{Birth/Death Proposal} \label{sec:birth move}
The birth/death proposal of Algorithm \ref{alg: MCMC across target configurations} initiates a new track or deletes an existing track and thus only generates proposed samples $(z'_{1:n},\mathbf{x}'_{1:n})$  (in step 2 which are then tested for acceptance) that move across dimension, i.e.\ is never intra-dimensional. Birth creates a track sequentially in time, until a stopping rule is met, by using the observed images to increase the probability of acceptance (we describe this sequential process in detail in Sec.\ \ref{sec:samplingbirth} below,) while death deletes a by choosing one at random.

The current MCMC input sample (Algorithm \ref{alg: MCMC across target configurations}) is $(z_{1:n},\mathbf{x}_{1:n})$ or $\{ (k,t_b^{k}, \mathbf{x}^{k})\}_{k=1}^K$ in the alternative parameterization of \ref{sec: Two equivalent notations for MTT}. Sampling from the  birth/death proposal commences by first choosing to create or delete a track with probability 0.5. If birth is chosen, a new track with birth time $t_b$ and states 
	$${\mathbf{x}}=[{x}_0,\ldots, {x}_{l-1}], \quad x_i=(a_i,s_i, v_i)$$ are proposed, where $(a_i,s_i, v_i)$ denote the intensity, spatial coordinates and velocity components of the state at time $t=t_b +i$. If death is chosen, one of the $K$ targets are randomly deleted, say target $k$ with probability $q_{d,\theta}(k|z_{1:n}, \mathbf{x}_{1:n}, y_{1:n})$. In the case of a birth, using $(z_{1:n}, \mathbf{x}_{1:n})$ and the newly created track, the ordering rule of Section \ref{sec: Two equivalent notations for MTT} is invoked to obtain the  MTT proposed sample $(z'_{1:n},\mathbf{x}'_{1:n})$. Assume the newly created target has label $k'$ in the alternative parameterization of $(z'_{1:n},\mathbf{x}'_{1:n})$. The acceptance probability is $\alpha_1=\min\{1,r_1\}$ where $r_{1}(z'_{1:n},\mathbf{x}'_{1:n}; z_{1:n},\mathbf{x}_{1:n})$ is
\begin{equation}
	\frac{p_{\theta}(z_{1:n}', \mathbf{x}'_{1:n}, y_{1:n})}{p_{\theta}(z_{1:n}, \mathbf{x}_{1:n},y_{1:n})}
	\frac{q_{d,\theta}(k'|z'_{1:n}, \mathbf{x}'_{1:n}, y_{1:n})}{q_{b, \theta}(t_b, \mathbf{x} |z_{1:n}, \mathbf{x}_{1:n}, y_{1:n}) }
	\label{eq: AcptRatio_BM_Img}
\end{equation}
The term $q_{b, \theta}(t_b, \mathbf{x} |z_{1:n}, \mathbf{x}_{1:n}, y_{1:n})$, defined in Sec.\ \ref{sec:samplingbirth} below, is the pdf of the newly created target states (which corresponds to term $Q(m' , u | m, x_m)$ in the denominator of \eqref{eq:TransMC}.) The term $q_{d,\theta}(k'|z'_{1:n}, \mathbf{x}'_{1:n}, y_{1:n})$ is the probability of deleting this newly created target.

If death is chosen, delete target $k$ of $\{ (i,t_b^{i}, \mathbf{x}^{i})\}_{i=1}^K$ with probability $q_{d,\theta}(k|z_{1:n}, \mathbf{x}_{1:n}, y_{1:n})$ and let  $(z'_{1:n},\mathbf{x}'_{1:n})$ be the new MTT state excluding target $k$. The acceptance probability is $\alpha_1=\min\{1,r_1\}$ where $r_{1}(z'_{1:n},\mathbf{x}'_{1:n}; z_{1:n},\mathbf{x}_{1:n})$ is 
\begin{equation}
	\frac{p_{\theta}(z_{1:n}', \mathbf{x}'_{1:n}, y_{1:n})}{p_{\theta}(z_{1:n}, \mathbf{x}_{1:n},y_{1:n})}
	\frac{q_{b, \theta}(t_b^k, \mathbf{x}^{k} |z'_{1:n}, \mathbf{x}'_{1:n}, y_{1:n}) }{q_{d,\theta}(k|z_{1:n}, \mathbf{x}_{1:n}, y_{1:n})}
	\label{eq: AcptRatio_DM}
\end{equation}

\subsection{Birth/Death Proposal: Creating a new target}\label{sec:samplingbirth}
The birth procedure below proposes a new track by using the \emph{residual} images to first construct the intensity and spatial coordinates of the track and then finally the velocity values of the track. For the current MCMC input sample (Algorithm \ref{alg: MCMC across target configurations}) $(z_{1:n},\mathbf{x}_{1:n})$, subtract the contribution of the $k_t^x$ targets and background intensity $b_t$ from the image $y_t$ to get the residual image $y_t^r$ where 
\begin{equation}
y_{t,i}^r=y_{t,i}-h_i(\mathbf{x}_t) - b_t, \qquad i=1,\ldots, m.
\label{eq:imageResidual}
\end{equation}
Match filter $y_t^r$ to get the image $y^f_t$ where the $j$-th pixel in the filtered image is  
\begin{equation} 
y^f_{t,j}=\frac{1}{E_{\bar{h}}}\sum_{i=1}^m y_{t,i}^r\bar{h}_i(j)
\label{eq:imagefiltered}
\end{equation}
where $\bar{h}_i(j)$ is defined in \eqref{Eq:PointObs}-\eqref{eq:psfandL} and $E_{\bar{h}}=\sum_{i=1}^m \bar{h}_i(j)^2$ is the energy of the filter $\{\bar{h}_i(j)\}_{i=1}^m$. (The sum that defines $y_{t,j}^f$  can be truncated to $i\in L(j)$.)
The rationale is that the presence of a target at or close to pixel $j$ will likely result in $y^f_{t,j}$ being a local maxima among pixels. Put another way, local maxima of $y^f_{t,j}$ are likely locations of targets.

\subsubsection{Proposing the initial state}
The move is commenced by choosing the birth time $t=t_b$ randomly from $1,\ldots,n$ and then followed by steps 1 and 2 below.
\paragraph*{Step 1}  Let \[G_t=\{1\leq i \leq m : y^f_{t,i} \textrm{ is a local maxima}, y^f_{t,i}\geq \gamma_t(\theta)\}\] and randomly  choose $i \in G_t$. $ \gamma_t(\theta)$ is a time-dependent threshold chosen to avoid peaks that are not likely to be target generated. A definition is given in the numerical section (Sec.\  \ref{sec: Numerical examples}.) 

 As a local intensity maxima is not necessarily target generated, perform a hypothesis test on the square of $l \times  l$ pixels $L(i)$ centered at chosen maxima $i \in G_t$. Let $y^r_{t,L(i)}=\{y^r_{t,j}, j \in L(i)\}$, $H_1$ the hypothesis that $y^r_{t,L(i)}$ is generated by a new born target and $H_0$ the converse that it is purely background noise generated. Calculate the test ratio
	\begin{align}
	  \rho(y_{t,L(i)}^r)
	=\frac{p(H_1)}{p(H_0)}\frac{p(y^r_{t,L(i)}|H_1)}{p(y^r_{t,L(i)}|H_0)} \label{eq:ModelRatio}
	\end{align}
While $p(y^r_{t,L(i)}|H_0):=\prod_{j\in L(i)} \mathcal{N}(y_{t,j}^r; 0,\sigma_{r,t}^2 )$ can be calculated analytically, $p(y^r_{t,L(i)}|H_1)$ is intractable but can be estimated, e.g. we use the Laplace approximation. (See Appendix~\ref{appdix: hypoTest} for details.)
The probability of $H_1$ is $p(H_1)=\sum_{k > k_b^t } \mathcal{P}(k,\lambda_b)$, which is the probability that the number of births exceeds $k_t^b$ where $k_t^b$ is the number of targets born at time $t$ in the current MCMC sample $z_{1:n}$. Set $p(H_0)=1-p(H_1)$. $H_1$ is accepted with probability $\min \{1, \rho(y_{t,L(i)}^r) \} $. Return (exit birth move) if $H_1$ is rejected. (Alternatively, $H_1$ could be accepted with probability $\rho /(1+\rho)$ which is less presumptuous than $\min \{1, \rho \}$.)

\paragraph*{Step 2} After accepting $H_1$, sample the intensity and position components of the initial target state by
	\[(A_0,S_0)\sim \mathcal{N}_{t,i}(\cdot),\] where $\mathcal{N}_{t,i}(\cdot)$ is a Gaussian derived from the Laplace approximation of the hypothesis test and is given in Appendix~\ref{appdix: hypoTest}. The subscript $(t,i)$ indicates this approximation is specific to pixels $L(i)$ of time $t$. 
	
\subsubsection{Proposing the remaining intensity-position trajectory}
The birth move continues  for $t=t_b+k$, $k>0$, using steps 1 and 2 above until a stopping rule is met. To conserve computations and increase its effectiveness, the range $1\leq i \leq m$ in the definition of $G_t$ is limited to a region of pixels $R_t$ where the next state would almost surely lie in. For example, $R_t$ can be determined by  the previous position $s_{t-1}$ and the upper limit on the target velocity.

The birth move stops at some time $t=t_d$, yielding a target lifespan $l=t_d-t_b$, when either $t>n$, $G_t$ is empty or $H_1$ is rejected. The output of this move is 
$(a_1,s_1),\ldots,(a_{l-1},s_{l-1})$ where $(a_0,s_0)$ was generated before.

\subsubsection{Proposing the velocity trajectory}
The output of the birth move thus far is $(a_0,s_0),\ldots,(a_{l-1},s_{l-1})$ which is the complete trajectory of intensity and positions of the new born target. The velocity components are now generated to yield
	$${\mathbf{x}}=[{x}_0,\ldots, {x}_{l-1}], \quad x_i=(a_i,s_i, v_i).$$
For linear Gaussian state dynamics, the velocity can be sampled conditioned on the spatial locations $s_0,\ldots,s_{l-1}$ and more generally, a Gaussian approximation/quadrature technique could be employed.\footnote{The numerical examples (synthetic and real data) assume Gaussian targets.}

\subsubsection{Proposal density of the birth move}  
Denoting $t_k=t_b+k$, $k=0,\ldots,l$, we can write the pdf for proposing  $(t_b, \mathbf{x})$ in the birth move as
\begin{align}
&q_{b,\theta}(t_b, \mathbf{x}| z_{1:n}, \mathbf{x}_{1:n},y_{1:n}) = q_0(t_b)q_1(a_0,s_0| y^r_{t_0}) \notag \\
& \times \prod_{k=1}^{l-1}q_2(a_k,s_k | a_{0:k-1},s_{0:k-1}, y^r_{t_k}) \notag\\
& \times q_3(\text{stop}|a_{0:l-1},s_{0:l-1}, y^r_{t_l})\;q_4(v_{0:l-1}|{a}_{0:l-1},{s}_{0:l-1})
\label{eq:BM}
\end{align}
where $q_0(t_b)$ is the probability of choosing the birth time  $t_b$, $q_{1}$ corresponds to proposing the target's initial intensity and position, and can be written as 
	\begin{align}
	&	q_{1}(a_0,s_0|y^r_{t_0})=\mathbb{I}[G_{t_0}\neq\emptyset] \nonumber \\
		& \times \sum_{i \in G_{t_0}}q(i |G_{t_0}) \min(1, \rho(y_{t_0,L(i)}^r)) \mathcal{N}_{t_0,i}(a_0,s_0)
\label{eq:birthinnitstate}
\end{align} where
       $ \mathbb{I}[G_{t_0}\neq\emptyset]$ is 1 if $G_{t_0}$ is non-empty; 
       $q(i|G_{t_0})$ is the probability of choosing local intensity peak  $i \in G_{t_0}$; 
       $\min(1, \rho(y_{t_0,L(i)}^r))$ is the probability of accepting hypothesis $H_1$ at peak $i \in G_{t_0}$; 
       $\mathcal{N}_{t_0,i}$ is the Gaussian density approximation for the initial value. 

The law $q_{2}$ is that for adding more states sequentially and may be written as
	\begin{align}
	&	q_2(a_k,s_k | a_{0:k-1},s_{0:k-1}, y^r_{t_k}) = \mathbb{I}[G_{t_k}\neq\emptyset]  \nonumber \\
	& \quad \times \sum_{i \in G_{t_k} } q(i | G_{t_k})  \min(1, \rho(y_{t_k,L(i)}^r)) \mathcal{N}_{t_k,i}(a_k,s_k )
	\label{eq:birthaddmorestates}
	\end{align}
which is  similar to  $q_{1}$  except that the previously created states $a_{0:k-1}$ and $s_{0:k-1}$ are used to calculate $\rho(y_{t_k,L(i)}^r)$, which is defined as in \eqref{eq:ModelRatio} with the difference here being $p(H_1)=p_s$ (target survival probability.)
The likelihood ratio term in \eqref{eq:ModelRatio} is derived in the Appendix, as is the Gaussian term $\mathcal{N}_{t_k,i}$ in \eqref{eq:birthaddmorestates}.

When $t_l=t_b+l > n$  stopping at $t_l$ is certain. Otherwise, the law $q_3$ corresponds to stopping due to the target not surviving, $G_{t_l}$ being empty, or the hypothesis test failing and is given by
	\begin{align*}
	&	q_3(\text{stop} | a_{0:l-1}, s_{0:l-1}, y^r_{t_l})\\
	&= 1 - p_s \mathbb{I}[G_{t_l}\neq\emptyset]  \sum_{i \in G_{t_l} } q(i | G_{t_l})  \min(1, \rho(y_{t_l,L(i)}^r))
	\end{align*}
For linear and Gaussian state dynamics, conditioned on $a_0,s_0,\ldots,a_{l-1},s_{l-1}$, the velocity can be sampled (exactly) since
$q_4$  will be a Gaussian distribution as well.

\subsection{Multi-step Extension/Reduction proposal}\label{sec:extendreduce}
This proposal extends or reduces the trajectory of a randomly chosen target. A target's trajectory is extended (/reduced) by bringing forward its birth (/death) time or delaying its death (/birth) time. The extra state values are then appended (/discarded) accordingly. Like the birth/death proposal, this proposal only moves the MCMC sample across dimension. 

The current MCMC input sample (Algorithm \ref{alg: MCMC across target configurations}) is $(z_{1:n},\mathbf{x}_{1:n})$ or $\{ (k,t_b^{k}, \mathbf{x}^{k})\}_{k=1}^K$ in the alternative parameterization of \ref{sec: Two equivalent notations for MTT}. Sampling from the multi-step extension/reduction proposal commences with choosing between extension and reduction equiprobably.
\paragraph*{Extension} From the subset of targets \textit{with} lifetimes less than $n$, randomly select a target and extension direction. (Without the lifetime restriction the chosen target cannot be extended further.) The direction of extension is
chosen equiprobably if both forward and backward extensions are permissible.
Assume target $k$ is selected for a forward extension, denoted $k_{+}$. A new (delayed) death time and trajectory extension  $\mathbf{x}=(x_1,x_2,\ldots)$ is proposed to yield the new extended trajectory $\hat{\mathbf{x}}^{k}=(\mathbf{x}^{k},\mathbf{x})$.
For a backward extension of target $k$, the event denoted by $k_{-}$, a new (earlier) birth time, denoted $\tau_b^{k}$ and trajectory extension  $\mathbf{x}$ is proposed to yield $\hat{\mathbf{x}}^{k}=(\mathbf{x}, \mathbf{x}^{k})$.
The ordering rule of Section \ref{sec: Two equivalent notations for MTT} is invoked to obtain the  MTT proposed sample $(z'_{1:n},\mathbf{x}'_{1:n})$ from the unaltered targets $\{ (t_b^{i}, \mathbf{x}^{i})\}_{i=1,i\neq k}^K$ and the extended target $( \tau_b^{k}, \hat{\mathbf{x}}^{k})$. The acceptance probability is $\alpha_2=\min \{1,r_2\}$ where $r_{2}(z'_{1:n},\mathbf{x}'_{1:n}; z_{1:n},\mathbf{x}_{1:n})$ is 
\begin{equation}
	\frac{p_{\theta}(z_{1:n}', \mathbf{x}'_{1:n}, y_{1:n})}{p_{\theta}(z_{1:n}, \mathbf{x}_{1:n}, y_{1:n})}
	\frac{q_{r,\theta}(z_{1:n},\mathbf{x}_{1:n}|z'_{1:n}, \mathbf{x}'_{1:n})}{q_{e, \theta}(k_{+/ -},{\mathbf{x}} |z_{1:n}, \mathbf{x}_{1:n}, y_{1:n})}.
	\label{eq:mstep_ext_red}
\end{equation}	
The probability density of choosing target $k$, extension direction and states ${\mathbf{x}}$
is denoted \\ $q_{e, \theta}(k_{+/ -},{\mathbf{x}} |z_{1:n}, \mathbf{x}_{1:n}, y_{1:n}) $ which is calculated similarly to the expression (\ref{eq:BM}) of the birth move and it is not repeated here. We  denote the total probability of making the return transition from 
$(z'_{1:n}, \mathbf{x}'_{1:n})$ to $(z_{1:n}, \mathbf{x}_{1:n})$ via the reduction, described next, with the term $q_{r,\theta}$ above. 

\paragraph*{Reduction}  Randomly select a target from the subset of targets with lifetimes exceeding one and then the reduction direction, either forwards or backwards, equiprobably. Let $k_{+}$ denote 
target $k$ for forward reduction and $k_{-}$ for backward reduction.
In the forward case, a reduction time point $t \in \{ t_b^k+1, \ldots, t_d^k -1 \}$ is chosen randomly and discard  the time $\{ t, \ldots, t_{d}^{k}-1 \}$ section of the track, causing $t$ to be the new death time. If backwards then $t \in \{ t_b^k, \ldots, t_d^k-2 \}$ is chosen randomly and the $\{ t_b^k, \ldots, t \}$ portion is discarded to yield a new birth time of $t+1$.  Let   $\tau_b^{k}$ denote the (possibly new) birth time, $\hat{\mathbf{x}}^{k}$ the retained trajectory and ${\mathbf{x}}$ the discarded forward/backward state trajectory of target $k$. The ordering rule of Section \ref{sec: Two equivalent notations for MTT} is invoked to obtain the  MTT proposed sample $(z'_{1:n},\mathbf{x}'_{1:n})$ from the the unaltered targets $\{ (t_b^{i}, \mathbf{x}^{i})\}_{i=1,i\neq k}^K$ and the reduced target $( \tau_b^{k}, \hat{\mathbf{x}}^{k})$.  Assume the reduced target has label $k'$ in the alternative parameterization of $(z'_{1:n}, \mathbf{x}'_{1:n})$.
Let $q_{r,\theta}(z'_{1:n}, \mathbf{x}'_{1:n}| z_{1:n},\mathbf{x}_{1:n})$  denote the total probability of making the transition from 
$(z_{1:n}, \mathbf{x}_{1:n})$ to $(z'_{1:n}, \mathbf{x}'_{1:n})$ via the described reduction step. The acceptance probability is $\alpha_2=\min \{1,r_2\}$ where 
$r_{2}(z'_{1:n},\mathbf{x}'_{1:n}; z_{1:n},\mathbf{x}_{1:n})$ is
\begin{equation}	
	\frac{p_{\theta}(z_{1:n}', \mathbf{x}'_{1:n}, y_{1:n})}{p_{\theta}(z_{1:n}, \mathbf{x}_{1:n}, y_{1:n})}
	\frac{q_{e, \theta}(k'_{+/ -},\mathbf{x} |z'_{1:n}, \mathbf{x}'_{1:n}, y_{1:n})}{q_{r,\theta}(z'_{1:n}, \mathbf{x}'_{1:n} | z_{1:n},\mathbf{x}_{1:n})}
	\label{eq:mstep_ext_red2}
\end{equation}	


\subsection{One-step Extension/Reduction proposal }

The intensity threshold $\gamma_t(\theta)$ used in the Birth (Sec.\ \ref{sec:birth move}))
and Extension (Sec.\ \ref{sec:extendreduce}) moves ignore the local
intensity peaks $G_{t}$ of the match filtered image $y_{t}^{f}$
that are below $\gamma_t(\theta).$ This may result in the next state
of a momentarily dim target not being detected. As a remedy, a new
\emph{one-step} extension/reduction proposal is defined. This proposal, which is to be regarded as a different proposal to multi-step extension/reduction (Sec.\ \ref{sec:extendreduce}), 
proceeds as multi-step except that it extends or truncates the 
trajectory of the selected target by one time point only. In particular, its acceptance probability is $\alpha_3=\min \{1,r_3 \}$ with $r_3$ defined as in  \eqref{eq:mstep_ext_red} and \eqref{eq:mstep_ext_red2} but with the following differences. The expression $q_{r,\theta}(z'_{1:n}, \mathbf{x}'_{1:n} | z_{1:n},\mathbf{x}_{1:n})$ in \eqref{eq:mstep_ext_red} (and \eqref{eq:mstep_ext_red2}) is now $1/2K$ (assuming $K$ targets have lifetimes exceeding one) since the probability of selecting a particular target is $1/K$ and then the reduction direction is $1/2$. Let the trajectory of the selected target prior to extension be $\mathbf{x}^{k}=(x_{0}^{k},\ldots,x_{l^{k}-1}^{k})$, then $q_{e,\theta}$ in \eqref{eq:mstep_ext_red} for $k_{+}$ is
\[
q_{e, \theta}(k_{+},{x} |z_{1:n}, \mathbf{x}_{1:n})= q_{e, \theta}(k_{+} | z_{1:n}, \mathbf{x}_{1:n}) f_{\psi}( x \vert x_{l^{k}-1}^{k})
\]
where the first factor is the probability of selecting $k_{+}$ (target $k$ and forward extension) and the extended state value is sampled from the prior model $f_{\psi}$.
For a backward extension or $k_{-}$, $q_{e, \theta}(k_{-},{x} |z_{1:n}, \mathbf{x}_{1:n})$ is
\[
q_{e, \theta}(k_{-} | z_{1:n}, \mathbf{x}_{1:n}) 
\frac{\mu_{\psi}(x)f_{\psi}(x_{0}^{k}\vert x)}{\int\mu_{\psi}(x')f_{\psi}(x_{0}^{k}\vert x')\mathrm{d}x'},
\]
the extended state is sampled from initial distribution $\mu_{\psi}$ of new targets conditioned on the value of its next state.

\subsection{State proposal}\label{sec:stateproposal}
\begin{figure}[h]
	\centering
	\includegraphics[width=.8\linewidth]{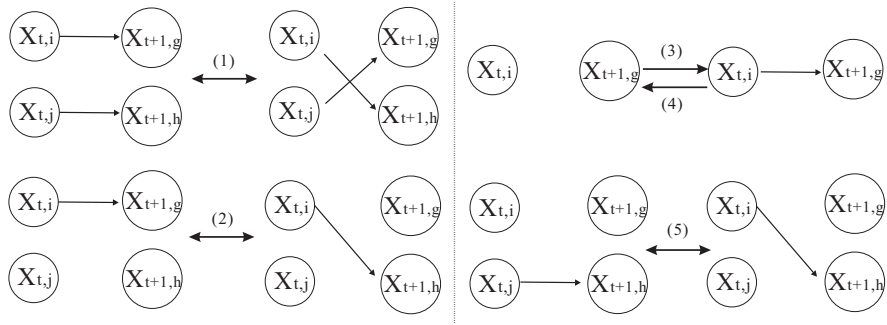}
	\caption{State move.}
	\label{fig:SM Img}
\end{figure}

This proposal chooses a pair of targets and swaps a section of their state trajectories. In particular, it randomly chooses a time $t<n$ and then randomly changes $I_{t+1}$, which is the vector that links targets in $\mathbf{X}_t$ and $\mathbf{X}_{t+1}$, as illustrated in Figure \ref{fig:SM Img}. 
When $X_{t,i}$ has descendant $X_{t+1,g}$,  it can propose to swap its descendant with that of $X_{t,j}$ (case $1$), or change its descendant to  the initial state $X_{t+1,h}$ of a target born at time $t + 1$ (case $2$), or to delete the link (case $4$). When $X_{t,i}$ has no descendant, it can be merged with a new-born target at time $t+1$ by linking to its initial state (case $3$), or steal another surviving target's descendant (case $5$). 
The state proposal is purely intra-dimensional (or in the context of Sec.\ \ref{sec:rjmcmcreview} it moves between models pairs $(m,m')$ satisfying $d_m=d_{m'}.$)

The current MCMC input sample (Algorithm \ref{alg: MCMC across target configurations}) is $(z_{1:n},\mathbf{x}_{1:n})$ or $\{ (k,t_b^{k}, \mathbf{x}^{k})\}_{k=1}^K$ in the alternative parameterization of \ref{sec: Two equivalent notations for MTT}. 
Sampling from the state proposal commences with choosing a time $t<n$ and a pair of targets (labels) $U=\{i, j \}$ from the total set of targets  $\{1,\ldots,K\}$ subject to targets $i$ and $j$ being alive at times $t$ and $t+1$ respectively. Choosing $i = j$ is permitted (then $U=\{ i \}$) and target $i$ must be alive at time $t$ and $t+1$. 
(For example, in implementation we chose a target $i$ from time $t$, and a target state value from the set of time $t+1$ targets with probability inversely proportional to the distance from target $i$'s state value at time $t$.)
We denote the probability of a particular selection by $q_{s,\theta} (t,U | \mathbf{x}_{1:n}, z_{1:n} )$.
If $i=j$ then split target $i$ into two targets as in case 4 of Figure \ref{fig:SM Img}. If $i\neq j$, pair the \emph{ancestors} of $i$ with the  \emph{descendants} of $j$ in the manner shown in Figure \ref{fig:SM Img} (all cases except 4), i.e. the swap alters trajectories  $\mathbf{x}^i$ and $\mathbf{x}^{j}$ to
\begin{eqnarray}
\mathbf{x}^i & \rightarrow & (x^i_0,\ldots,x^i_s, x^{j}_{s'},\ldots,x^{j}_{l^{j}-1}) =: \hat{\mathbf{x}}^{i} \nonumber \\
\mathbf{x}^{j} & \rightarrow & (x^{j}_0,\ldots,x^{j}_s, x^{i}_{s'}, \ldots,x^{i}_{l^{i}-1})  =: \hat{\mathbf{x}}^{j}  \label{eq:swaptraj}
\end{eqnarray}
where $t_b^{i}+s=t$, $t_b^{j}+s'=t+1$. Note the birth time (of at most one target) may change. Call the birth times after the swap $\tau_b^i$, $\tau_b^j$

 The next part of the state move then proposes a change to the continuous variables of the affected targets $k\in U=\{i,j\}$ form $\hat{\mathbf{x}}^{k}$ to $\tilde{\mathbf{x}}^{k}$ to increase the chance of the move being accepted. This is because changing the links between targets at time $t$ and $t+1$ can cause a mismatch in the velocity and intensity of the newly formed links. As such, the state move will then propose a change to the velocity and intensity components of the affected targets while retaining their original spatial position components. Let $q_{s,\theta}(t,U,(\tilde{\mathbf{x}}^{k})_{k\in U} |\mathbf{x}_{1:n}, z_{1:n},y_{1:n} )$ denote the joint pdf of selecting $(t,U)$ and the change $(\hat{\mathbf{x}}^{i},\hat{\mathbf{x}}^{j}) \rightarrow (\tilde{\mathbf{x}}^{i},\tilde{\mathbf{x}}^{j})$. (See Appendix \ref{app:state} for the expression.)
 Finally, we let $(z'_{1:n}, \mathbf{x}'_{1:n})$ denote MTT state (in the original parameterization) of the unaltered $\{ t_b^{k}, \mathbf{x}^{k}\}_{k\in \{1\ldots,K\}/U }$ and altered targets $\{ \tau_b^{k}, \tilde{\mathbf{x}}^{k}\}_{k\in U }$. The acceptance probability of the state move is then $\alpha_3=\min\{1,r_3\}$ where $r_{3}(z'_{1:n},\mathbf{x}'_{1:n}; z_{1:n},\mathbf{x}_{1:n})$ is
\begin{equation}
	\frac{p_{\theta}(z_{1:n}', \mathbf{x}'_{1:n}, y_{1:n})}{p_{\theta}(z_{1:n}, \mathbf{x}_{1:n}, y_{1:n})}
	\frac	{q_{s,\theta} (t,U', ({\mathbf{x}}^{k})_{k \in U'} | \mathbf{x}'_{1:n}, z'_{1:n}, y_{1:n} )}
	{q_{s,\theta} (t,U, (\tilde{\mathbf{x}}^{k})_{k \in U} | \mathbf{x}_{1:n}, z_{1:n}, y_{1:n} )} \label{eq: AcptRatio_SM_Img}
\end{equation}	
$U'$ is the label set of the targets with swapped components in the alternate parameterisation of $(z'_{1:n}, \mathbf{x}'_{1:n})$.

\section{Numerical examples}\label{sec: Numerical examples}
This section presents the two main numerical examples. The first one uses synthetic data and assumes known MTT parameters so that a fair comparison can be made between our MCMC tracker (Algortihm~\ref{alg: MCMC for static parameter estimation} excluding parameter learning) and  
 the multi-Bernoulli (MB) filter of \citep{vo2010joint}. (The MB tracker does not learn parameters.) The numerical results  will  demonstrate the performance improvements of our method when tracking targets that are close to each other with overlapping illumination regions. The second example is a real-data example that applies  Algorithm~\ref{alg: MCMC for static parameter estimation} to track  Fab\footnote{ Fab (Fragment antigen-binding) is a region on an antibody that binds to antigens.} labelled Jurkat T-cells. The tracking method currently used by biochemists \citep{weimann2013quantitative} extracts point measurements from the images and then connects them to form trajectories using a nearest neighbour type  method. Our algorithm will be  shown to outperform theirs when tracking dim targets as well as targets with overlapping illumination regions. 
All simulations were run in Matlab on a PC with an Intel i5 $2.8$ GHZ $\times2$ processor. 
 
Recall the definition of an individual target's state $X_t=(A_t,S_t(1),S_t(2),V_t(1),V_t(2))$ in \eqref{eq:state}. For the numerical examples, we use a drifting intensity and near constant velocity motion model,
\begin{align*}
A_t &= A_{t-1}+ U_{t},\quad U_{t} \overset{\text{i.i.d.}}{\sim} \mathcal{N}(0,\sigma_i^2), \\
[S_t(j),V_t(j)]&=[S_{t-1}(j)+\delta V_{t-1}(j),V_{t-1}(j)]+U_{t,j},\\
&\quad U_{t,1}^\textrm{T} \overset{\text{i.i.d.}}{\sim} \mathcal{N}(0,\sigma_x^2 \Sigma), 
U_{t,2}^\textrm{T} \overset{\text{i.i.d.}}{\sim} \mathcal{N}(0,\sigma_y^2 \Sigma), 
\end{align*}
where $\delta$ is the (known) sampling interval and 
\begin{equation*}
	\Sigma \!=\! \begin{pmatrix} 
	\delta^{3}/{3} & \delta^{2}/{2} \\ \delta^{2}/{2} & \delta \end{pmatrix}.
\end{equation*}
	The initial hidden state is assumed to be Gaussian  distributed with mean $\mu_{b} = (\mu_{bi},\mu_{bx},\mu_{by},0,0)^\textrm{T}$ and covariance $\Sigma_{b}=\text{diag}(\sigma_{bi}^2, \sigma_{bp}^2, \sigma_{bp}^2, \sigma_{bv}^2,\sigma_{bv}^2)$. The mean of the initial velocity is set to be $0$ in the absence of more information but this still can yield directional motion if the observations support this. All the parameters $\psi$ of the hidden state dynamics are (see \eqref{eq:state})
	\[
	\psi=
	( \mu_{bi},\mu_{bx}, \mu_{by}, \sigma_{bi}^2, \sigma_{bp}^2, \sigma_{bv}^2, \sigma_i^2, \sigma_x^2, \sigma_y^2).
	\]
and augmenting $\psi$ with the parameters of the target birth/death and observation models gives
\[
\theta=(\psi, p_s, \lambda_b, b_1, \sigma_{r,1}^2,\ldots,b_n, \sigma_{r,n}^2).
\]
\emph{Prior for $\theta$:} All the variance components above have independent priors, which is the inverse gamma distribution $\mathcal{IG}(\alpha_0, \beta_0)$ with (common) shape $\alpha_{0}$ and scale $\beta_{0}$ parameters.
(Setting $\alpha_0 \ll 1$ and $\beta_0 \ll 1$ yields a less informative prior.) Given $\sigma_{bi}^2$, $\sigma_{bpx}^2$, $\sigma_{bpy}^2$ and $\sigma_{r,t}^2$ (for $t=1,\ldots,n$), the priors of $\mu_{bi}$, $\mu_{bx}$, $\mu_{by}$ and $b_t$ are
$\mu_{bi}|\sigma_{bi}^2\sim \mathcal{N}(\mu_0,\sigma_{bi}^2/n_0)$, 
$\mu_{bx}|\sigma_{bp}^2\sim \mathcal{N}(\mu_0,\sigma_{bp}^2/n_0)$, 
$\mu_{by}|\sigma_{bp}^2\sim \mathcal{N}(\mu_0,\sigma_{bp}^2/n_0)$, 
$b_t|\sigma_{r,t}^2\sim \mathcal{N}(\mu_0,\sigma_{r,t}^2/n_0)$. 
These are made  more diffused by setting  $n_0$ and $\mu_0$ to be small. 
The conjugate priors of $p_{s}, \lambda_{b}$ are 
\[
p_s\sim \text{Unif} (0,1),\quad \lambda_b \sim\mathcal{G}(\alpha_0,\beta_0),
\]
where $\text{Unif}(a,b)$ and $\mathcal{G}(\alpha,\beta)$ represent (respectively) the uniform distribution over $(a,b)$ and the gamma distribution with shape parameter $\alpha$ and scale parameter $\beta$. We set $\alpha_{0} \ll 1, \beta_{0} \gg 1$ to make the prior less informative. (If prior knowledge is available, a more appropriate choice of $(\alpha_{0}, \beta_{0})$ can be made.) $K_{t}^{s}$ is a Binomial r.v.\ with success probability $p_{s}$  and number of trials $K_{t-1}^{x}$. $K_t^b$ is a Poisson r.v.\ with rate $\lambda_b$. Thus their posteriors are  
\begin{align*}
p_{s} | z_{1:n},y_{1:n} &\sim \mathcal{B} \biggl(1+\sum _{t=1}^n k_t^s,\, 1+\sum_{t=2}^n (k_{t-1}^x-k_t^s)\biggr),\\
 \lambda_{b} | z_{1:n}, y_{1:n} &\sim \mathcal{G} \biggl(\alpha_{0} + \sum_{t=1}^{n} k_{t}^{b},\, (\beta_{0}^{-1} + n)^{-1} \biggr).
 \end{align*}

The illuminated region  $L(s)$ is   an $l \times l$ square region of pixels centered at $s$, with $l=1+\lceil4\sigma_h\rceil / \Delta$ 
 where $\lceil\cdot\rceil$  rounds up its argument. The intensity threshold  $\gamma_t(\theta)$ is chosen to be $\gamma_t(\theta)=\min(\mu_{bi}-3\sigma_{bi}, 3\sigma_{r,t} / \sqrt{E_{\bar{h}}})$ using the following rationale.
We expect $y^f_{t,j}$ in \eqref{eq:imagefiltered} to exceed $\mu_{bi}-3\sigma_{b,i}$ (mean birth illumination minus 3 times its standard deviation) with high probability if a target is present in pixel $j$ of the residual image in \eqref{eq:imageResidual}. 
However, assuming no targets illuminate pixels $L(j)$
of the residual image, $\sigma_{r,t} / \sqrt{E_{\bar{h}}}$ is the standard deviation of $y^f_{t,j}$. With high probability,
$y^f_{t,j}$ should not be exceed $3\sigma_{r,t} / \sqrt{E_{\bar{h}}}$ and avoids triggering detection.

\subsection{Comparison with the  multi-Bernoulli tracker}\label{subsec:MB}
We compared our algorithm with the MB tracker \citep{vo2010joint}.  
Unlike the subsequent real data example, this synthetic case assumed $b_t=0$ and $\sigma_{r,t}^2=\sigma_r^2$ for all $t$ (see \eqref{eq:mttobs}.) 
 We synthesised  $50$ frames (images) of $168 \times 184$ pixels each using the parameter vector
\begin{equation}
\psi^*=(30, 0, 0, 4, 25, 3, 0.5, 0.3, 0.7), \theta^*=(\psi^*,0.95,0.3,1).
\label{truetheta}
\end{equation} 

\begin{figure}
	\begin{subfigure}[MCMC tracker]{.49\linewidth}
		\centering
		\includegraphics[width=.99\linewidth, height=1.1\linewidth]{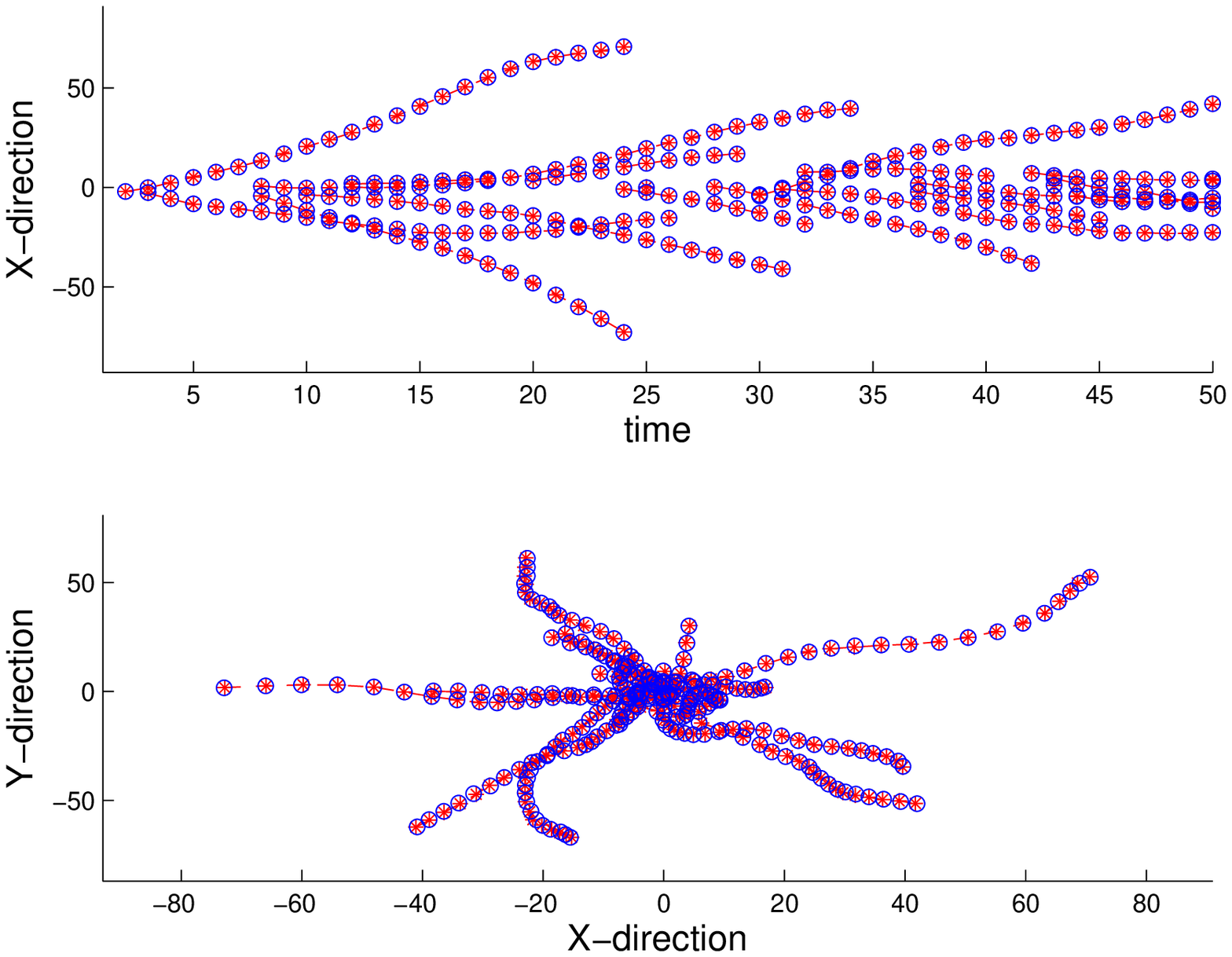}
		\caption{MCMC tracker}
		\label{fig:MCMC_tracker}
	\end{subfigure}
	\begin{subfigure}[multi-Bernoulli tracker]{.49\linewidth}
		\centering
		\includegraphics[width=.99\linewidth,height=1.1\linewidth]{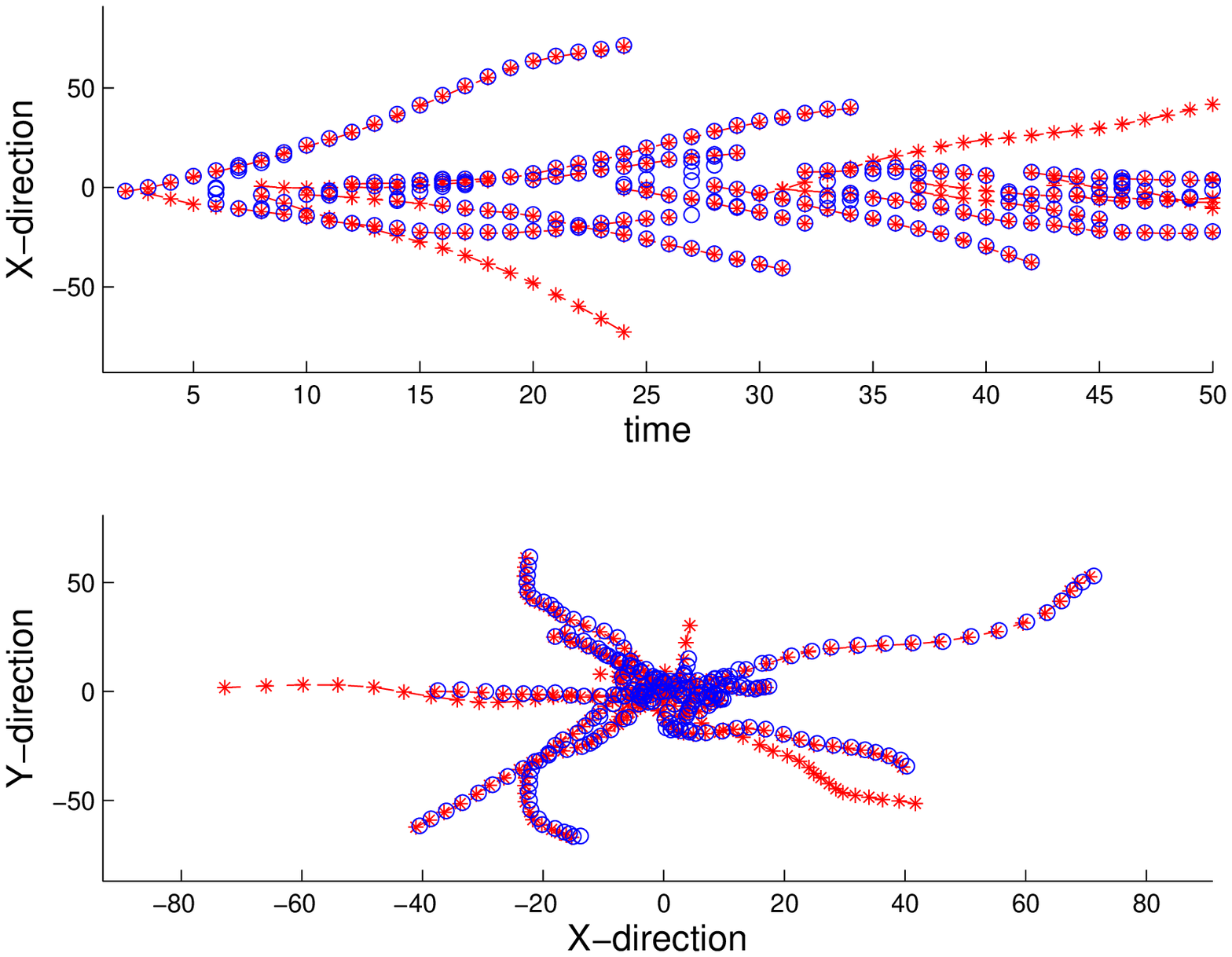}
		\caption{multi-Bernoulli tracker}
		\label{fig:MB_tracker}
	\end{subfigure}
	\caption{Comparison with the multi-Bernoulli filter in \cite{vo2010joint}.  Tracks labelled $-\ast-$ (red) is ground truth while (blue) circles are the estimates. 
}
\end{figure}

We set  $\sigma_h^2=1$, $\Delta = 1$.  This gives a $5 \times 5$ pixel square for the illuminated region $L(s)$ (see (\ref{Eq:PointObs})). 
From \eqref{Eq:PointObs}, define  
$\text{SNR}=20\log(\frac{a \Delta^2/2\pi \sigma_h^2}{\sigma_r}).$
For $a=\mu_{bi}=30$, the initial SNR is $13.6$ dB.
The synthetic data had  $20$ targets whose trajectories are shown in  Figure \ref{fig:MCMC_tracker} along with the trajectories  obtained by running the MCMC tracker with $n_1=30$, $n_2=1$ (and $n_3=0$) 
and $15$ particles per target for the particle Gibbs sampler. 
Figure \ref{fig:MCMC_tracker} shows all targets being tracked completely. In contrast, Figure \ref{fig:MB_tracker} shows the output of the MB tracker of \cite{vo2010joint}. The birth process assumed by the MB tracker has   four terms each of which  has the same initial distribution $\mathcal{N}(\cdot| \mu_b, \Sigma_b)$  and existence probability $0.1$.  Pruning and merging targets are performed as suggested in \cite{vo2010joint} to eliminate tracks with existence probability less than $0.01$ and merge two tracks when they fall within a fraction ($3/4$) of a pixel size in distance. The number of particles assigned  for each hypothesised target in MB tracker is restricted between $5000$ and $8000$. In Figure \ref{fig:MB_tracker}, it is seen that some tracks are lost after they cross, which is the main limitation of  the MB tracker as pointed out in \cite{vo2010joint}. This is because crossing targets invalidates the crucial assumption, necessary to derive the MB tracker, that the illuminated  region of the targets do not overlap. In terms of the computation time, the MB tracker take less than one minute to run while the MCMC tracker takes $6$ minutes. Closely spaced targets are common place in many applications, an example being the real-data experiment reported below. Our MCMC tracker should be viewed as a method applicable to closely spaced targets and not as a competitor of a technique optimized for non-overlapping targets like the MB tracker.

\begin{figure}
	\begin{subfigure}{.48\linewidth}
		\includegraphics[width=0.99\linewidth, height=0.65\linewidth]{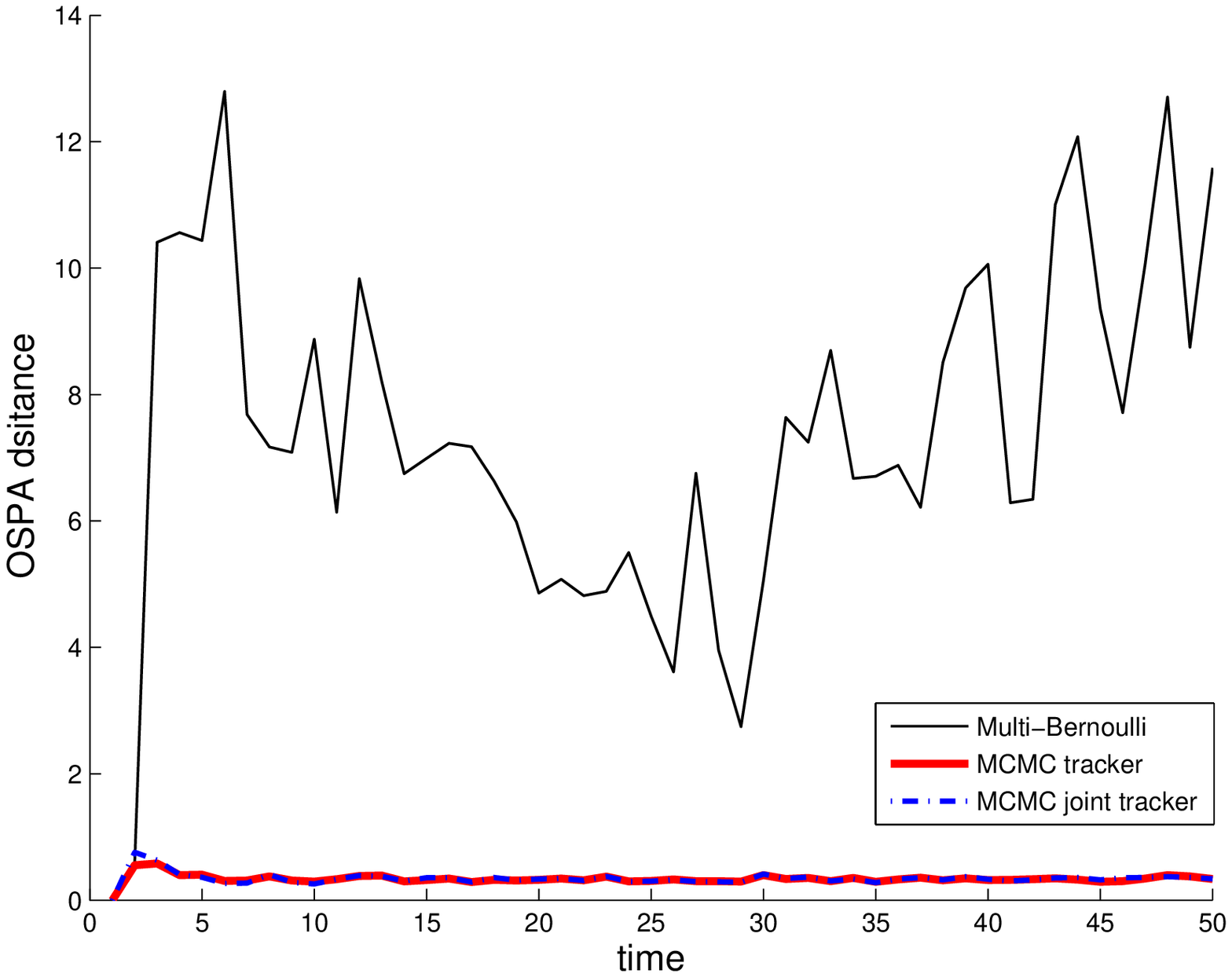}
		\caption{OSPA}
		\label{fig:OSPA}
	\end{subfigure}
	\begin{subfigure}{.48\linewidth}
		\centering
		\includegraphics[width=.99\linewidth, height=0.65\linewidth]{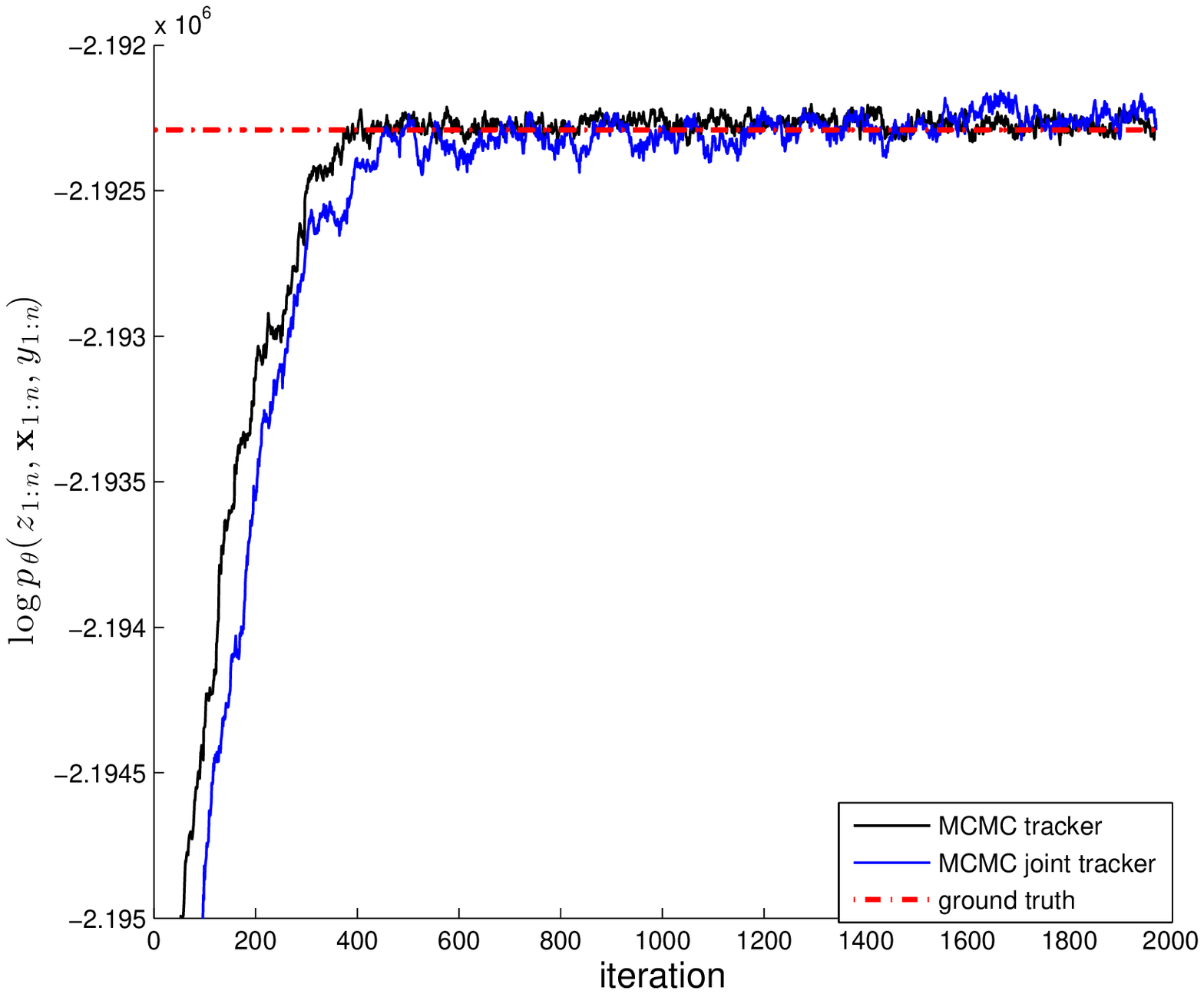}
		\caption{Log density}
		\label{fig:logLP}
	\end{subfigure}
	\caption{{\bf(a)} Comparing OSPA tracking error of the multi-Bernoulli filter (top solid black line) with Alg.\ \ref{alg: MCMC for static parameter estimation} with and without parameter learning (lower traces.) OSPA error of MCMC almost equal when $\theta^\ast$ is known or learnt during tracking. {\bf (b)} Plot  of $p_{\theta*}(z^{(i)}_{1:n},\mathbf{x}^{(i)}_{1:n}, y_{1:n})$ (Alg.\ \ref{alg: MCMC for static parameter estimation} with known $\theta^*$) and $p_{\theta^{(i)}}(z^{(i)}_{1:n},\mathbf{x}^{(i)}_{1:n}, y_{1:n})$ (Alg.\ \ref{alg: MCMC for static parameter estimation} with parameter learning) against MCMC iteration $i$. Horizontal (red) line indicates ground truth $p_{\theta^{*}}(z^{*}_{1:n},\mathbf{x}^{*}_{1:n}, y_{1:n})$.}
	\end{figure}

The previous comparison was done assuming known model parameters. The current example revisits the same data set assuming $\theta^\ast$ in (\ref{truetheta}) is unknown to Algorithm~\ref{alg: MCMC for static parameter estimation}, which was re-run with the initial parameter set to \[\theta^{(0)}=(45,10, 5, 
8, 50, 6,
 3,1, 1.5,
  0.6, 1, 4),\] 
  while MB output of the previous example was used.
 (The MB was given the true parameter as it does not incorporated parameter learning. The MCMC outputs of  Algorithm~\ref{alg: MCMC for static parameter estimation} will be denoted $(z^{(i)}_{1:n},\mathbf{x}^{(i)}_{1:n},\theta^{(i)})$.) OSPA distances  \citep{vo2010joint} of the three algorithms are plotted in Figure \ref{fig:OSPA}. 
Figure \ref{fig:OSPA} shows that the tracking performance with unknown parameters is similar to the known case reported earlier. A further verification is the probability density values plotted in Figure \ref{fig:logLP} where $p_{\theta*}(z^{(i)}_{1:n},\mathbf{x}^{(i)}_{1:n}, y_{1:n})$ was calculated from the previous experiment (Algorithm \ref{alg: MCMC for static parameter estimation} with known parameters) and $p_{\theta^{(i)}}(z^{(i)}_{1:n},\mathbf{x}^{(i)}_{1:n}, y_{1:n})$ are the density values from Algorithm~\ref{alg: MCMC for static parameter estimation} with parameter learning.  

\begin{figure}
	\centering
	\includegraphics[width=\linewidth,height=.7\linewidth]{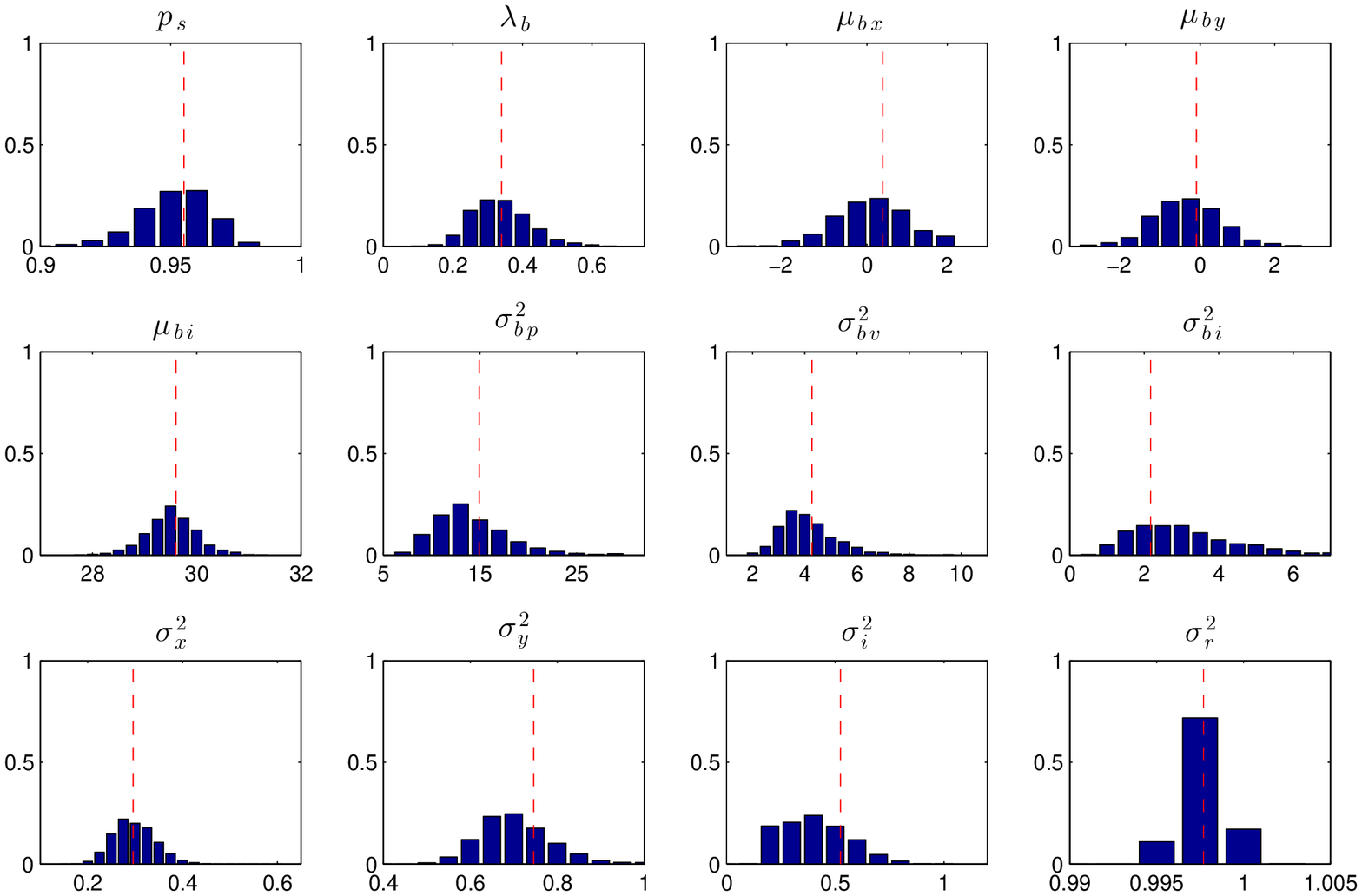}
	\caption{Histograms of $2000$ parameter samples obtained by  Algorithm~\ref{alg: MCMC for static parameter estimation}. Vertical (red) lines indicate the MLE estimate in \eqref{eq: theta_star_Img}.}
	\label{fig:Para_syth_hist}
\end{figure}

Figure \ref{fig:Para_syth_hist} shows the histograms of $2000$ post burn-in parameter samples of Algorithm~\ref{alg: MCMC for static parameter estimation} as the approximation of $p(\theta | y_{1:n})$. The (red) dashed lines show the MLE estimate $\hat{\theta}$ obtained using the true value of the latent variables, i.e. $(z^{\ast}_{1:n}, \mathbf{x}_{1:n}^\ast)$. Specifically, $\hat{\theta}$ is comprised of (see \eqref{eq: density of y given x,z}, \eqref{eq: density of z}, \eqref{eq: density of x given z}) 
\begin{align}
(\hat{p}_s,\hat{\lambda}_b) &= \arg \max_{p_s,\lambda_b} p(z^*_{1:n}), \quad
\hat{\psi}= \arg \max_{\psi} p(\mathbf{x}^*_{1:n}|z_{1:n}^*),\nonumber \\
\hat{\sigma}_r &= \arg \max_{\sigma_r} p(y_{1:n}|\mathbf{x}^*_{1:n},z_{1:n}^*).
\label{eq: theta_star_Img}
\end{align}
As a correctness check, for an uninformative prior,  the posterior modes should be consistent with MLE of $\theta^\ast$. The true MLE is $\arg \max_{\theta} p_{\theta}(y_{1:n})$, which will be different from (\ref{truetheta}), is not available as 
$p_{\theta}(y_{1:n})$ is intractable. We use $\hat{\theta}$ of \eqref{eq: theta_star_Img} as the surrogate. Note the modes of the posterior do coincide with the surrogate MLE.

\subsection{Experiments on Fab labelled Jurkat T-cells}\label{sec:cell}
The source of the data were Jurkat T-cells, an immortalised cell line of human T-lymphocytes which plays an important function in immune response. Cells were imaged using a microscopy technique called \emph{total internal reflection fluorescence microscopy}. The cell's molecules of interest were bound to antibodies labelled with a bright green-fluorescent dye (Alexafluo48). It is known that the number of labeled molecules (or `targets') can be high; at physiological levels there can be several molecules per square micron.

The data is comprised of  $20$ frames of  $115 \times 120$ pixel images with a pixel size $176nm$ and a frame rate of $17.8$ frame/s. The diffusion coefficient $D$ is expected to be in the region of  $0.01$-$0.1 um^2/s$. 
This implies the displacement of molecules between consecutive frames is expected to be  in the order of $0.1$-$1$ pixel. The estimated initial SNR here is around $10$dB (estimated from the tracking result). Figure~\ref{fig:Img_comp} (left) shows one frame of the observed  images.

\subsubsection{Parameter initialisation} In choosing the initial parameter vector $\theta^{(0)}$, some components were chosen arbitrarily while others were guided by the observed images. (Note that  the initialisation step does not need to be overly precise as our algorithm does not depend on specific initial values to work.) We set $p_s=0.6, \lambda_b=0.2$ arbitrarily; $\mu_{bx}^{(0)}=\mu_{by}^{(0)}=60$ to coincide with the image centres  since the centres appear much brighter than the periphery;  $\mu_{bi}^{(0)}=70$ is calculated from \eqref{Eq:PointObs} 
assuming a target is in the middle of the brightest pixel of the first frame;  set $(\sigma_{bp}^2 )^{(0)} = 400$ by roughly observing that the 
bright spots are sparse and span the whole image; $(\sigma^{2}_{bv})^{(0)}=1$ covers the velocity range of the molecules ($0.1$-$1$ pixel per image) and flat enough to allow different possible  diffusion coefficients; $(\sigma_{bi}^{2})^{(0)}=100$ arbitrarily; $(\sigma^{2}_{x})^{(0)}=(\sigma^{2}_{y})^{(0)}=0.1$ for small initial driving state noise; $(\sigma^{2}_{i})^{(0)}=25$ arbitrarily. The time-varying observation noise statistics, mean $(b_t)^{(0)}$ and variance $(\sigma_{r,t}^{2})^{(0)}$, are initialised to equal the mean and variance of the pixel intensities at that time since bright pixels are sparse in the images.

The point spread parameter $\sigma_h$ is not estimated and fixed at $\sigma_h=2$ (normalised with $\Delta^2$). The illuminated region $L(s)$ has $9\times 9$ pixels. The value of $\sigma_h$ if often known to the experimentalist otherwise, a bit tuning is required.\footnote{When $\sigma_h$ is too small,  the illuminated region taken into account is smaller than it should be which would cause  many more targets than expected. In that case, we should increase $\sigma_h$.} It could also be estimated as part of $\theta$.

\subsubsection{Comparison with \cite{weimann2013quantitative}}
\begin{figure*}
	\begin{subfigure}{.45\linewidth}
		\centering
		\includegraphics[width=\linewidth]{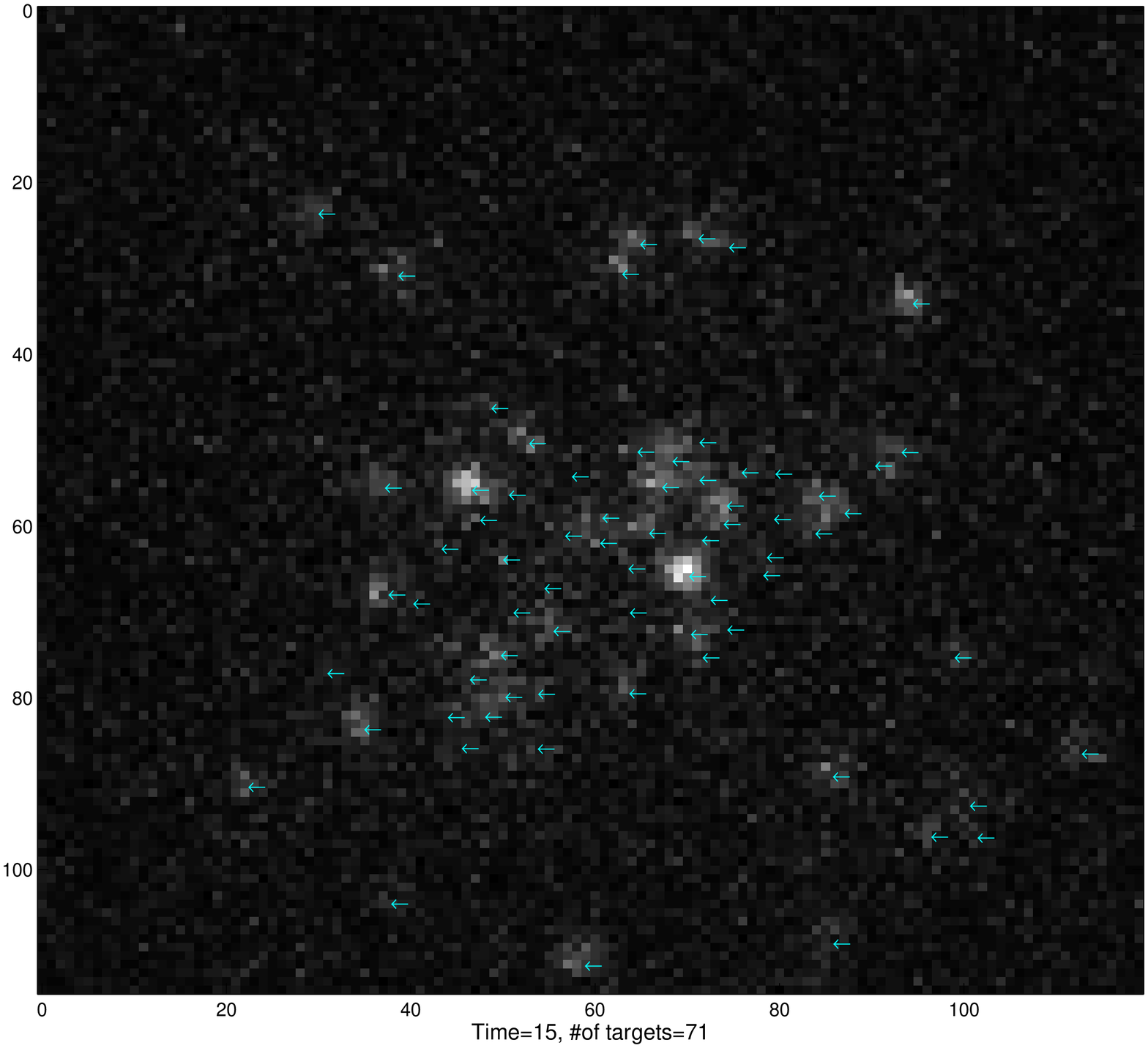}
	\end{subfigure}
	\begin{subfigure}{.45\linewidth}
		\centering
		\includegraphics[width=\linewidth]{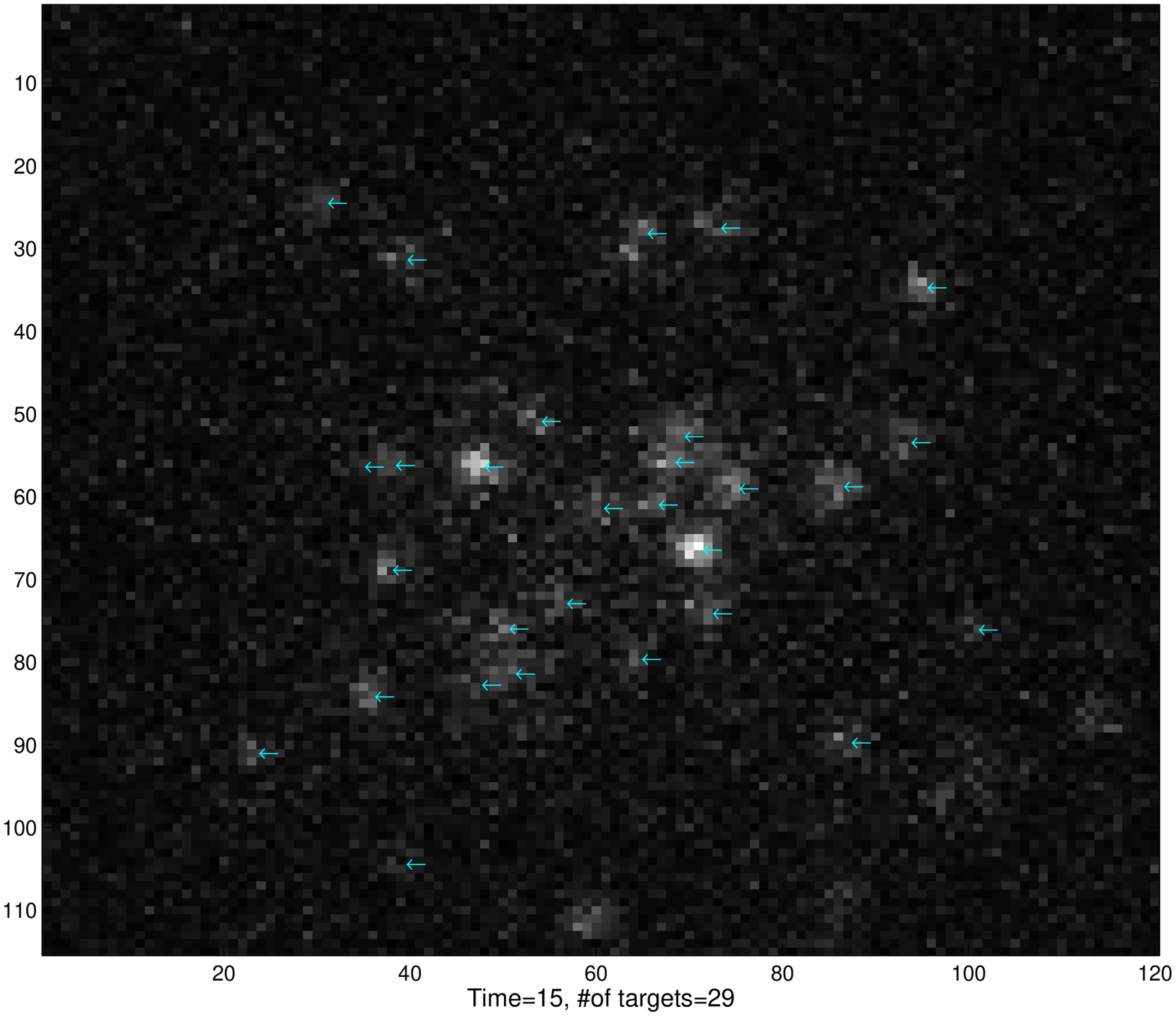}
	\end{subfigure}
	
	 \begin{subfigure}{.45\linewidth}
		\centering
		\includegraphics[width=0.99\linewidth]{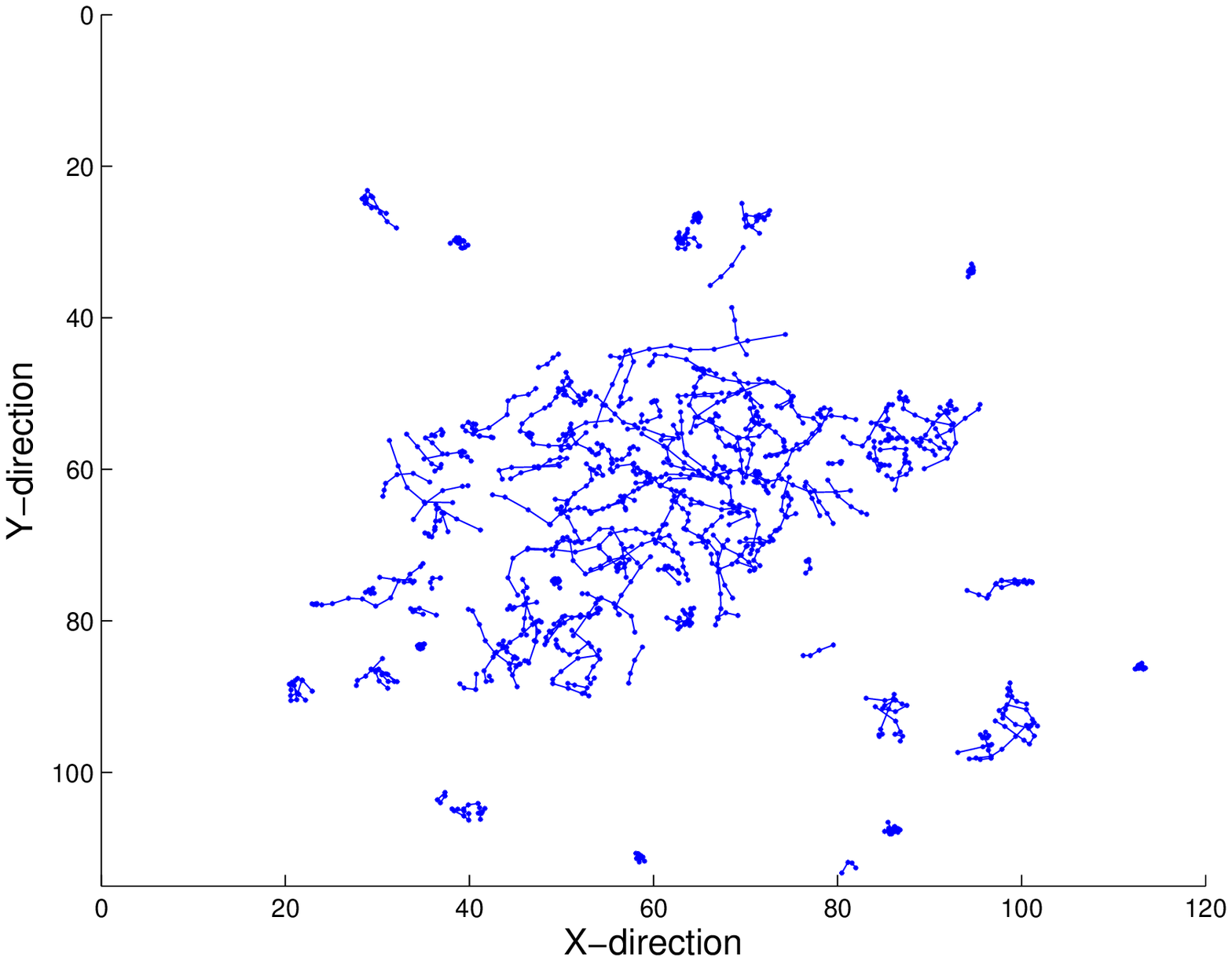}
		\label{fig:track_chem_Lan}
	\end{subfigure}
	\begin{subfigure}{.45\linewidth}
		\centering
		\includegraphics[width=0.99\linewidth]{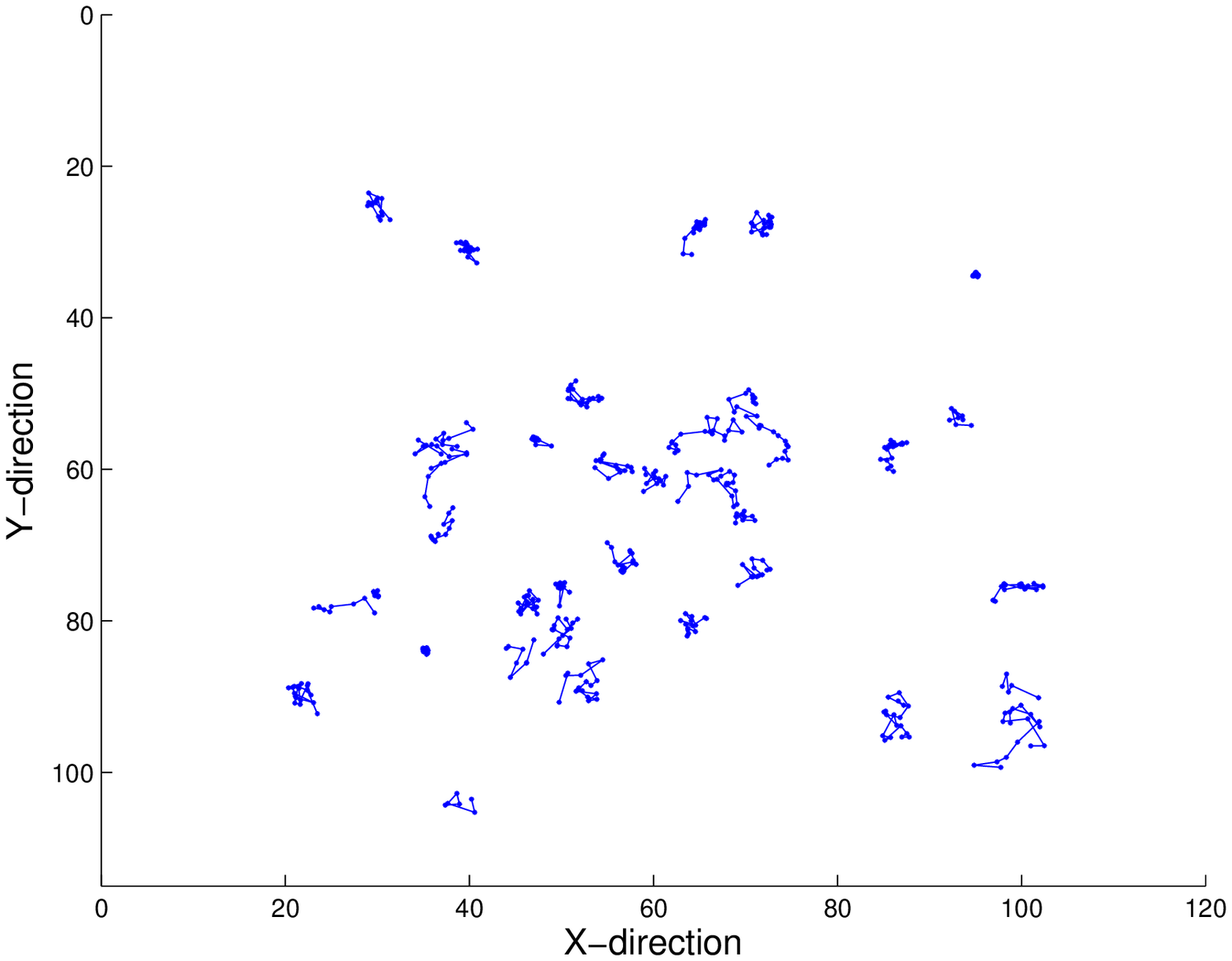}
		\label{fig:track_chem}
	\end{subfigure}

	\caption{Estimated target positions in frame $t=15$ of a video of fluorophore labelled molecules diffusing in the cell membrane of a Jurkat-T cell indicated by the (cyan) arrows. Top-left gives the estimated positions derived from one (representative) sample of the tracking result of Algorithm~\ref{alg: MCMC for static parameter estimation}, top-right for the method  in \cite{weimann2013quantitative}. Bottom-left is the tracking result of our method, bottom-right the method  in \cite{weimann2013quantitative}.}
	\label{fig:Img_time6}
\end{figure*}

\begin{figure}
	\centering
	\includegraphics[width=\linewidth]{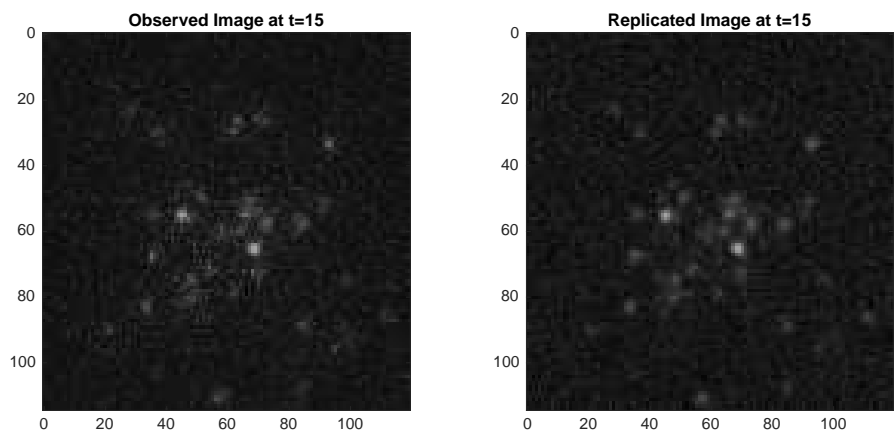}
	\caption{Comparison of the observed image (left) and the  replicated one (right) at frame $t=15$ for the real-data experiment.}
	\label{fig:Img_comp}
\end{figure}

The tracking algorithm of \cite{weimann2013quantitative} is a nearest-neighbour method which has an image pre-processing step to extract point measurements. Tracks are then  created by  connecting point  measurements nearest to each other.  A set of  consecutive frames are considered at the same time to allow temporary mis-detections. (See \cite{weimann2013quantitative}  for more details.)  One of the main disadvantages of this heuristic method is that  it may miss  targets moving close to each other with overlapping illumination regions, as only one point measurement may be extracted from a comparably big bright region; another disadvantage is  the user-defined hard-threshold which may cause the targets with lower intensities completely missed. Take frame $15$ in Figure~\ref{fig:Img_time6} as an  example. Algorithm~\ref{alg: MCMC for static parameter estimation} detects more targets in the centre  and  marginal regions of the frame, compared to the  method in \cite{weimann2013quantitative}.  (\cite{weimann2013quantitative} mentions indeed targets were missed in the marginal regions.) Figure~\ref{fig:Img_time6} compares the tracked positions of the molecules of these two algorithms. As a verification of our result, in the absence of ground truth, we compare in Figure  \ref{fig:Img_comp} the true and synthesised image (based on our estimated tracks and model parameters) at frame $t=15$. Their likeness offers some reassurance.

\section{Conclusion}
We have proposed a new MCMC based MTT  algorithm for joint tracking and parameter learning that works directly with image data and avoids the need to pre-process to extract point observations.  In numerical examples, we demonstrated improved performance in difficult tracking scenarios involving many targets with overlapping illumination regions, over competing methods \cite{weimann2013quantitative, vo2010joint}, which was achieved by targeting the exact posterior using MCMC.  We do not advocate that our MCMC technique should replace an online method or optimised methods for non-overlapping targets such \cite{vo2010joint}. It is an alternative that works without major underlying limiting assumptions (like non-overlapping) and can be used to refine online estimates. Possible future works include the design of more efficient proposals for our MCMC routine, parellelization and performance optimization for very high density  tracking.

\appendix
\section{Hypothesis testing}\label{appdix: hypoTest}
 The term of \eqref{eq:ModelRatio} to be calculated is 
 $p( y^r_{t,L(i)}  | H_1)$ where $y^r_{t,L(i)}=\{y^r_{t,j}\}_{j \in L(i)} $ and $i \in G_t$ is a pixel that is a local maximum of $y_t^f$. Let $(r,c)$ be its row and column number and
 \[(\bar{a},\bar{s})=(y^f_{t,i},\Delta r, \Delta c).\]    Vector 
 $(\bar{a},\bar{s})$ can be interpreted as the likely intensity $\bar{a}$ and location $\bar{s}$ of an undetected target.
 Below we just write $L$  as the set of pixels under consideration instead of $L(i)$.
 
 Let $x=(a,s,v)\in \mathbb{R} \times \mathbb{R}^2 \times \mathbb{R}^2$ where as before $a$ denotes intensity, $s=(s(1),s(2))$ spatial coordinates and $v=(v(1),v(2))$ spatial velocity.
(Recall that a pixel illumination is not a function of velocity.) The aim is to calculate
 \begin{align}
 p(y^r_{t,L}|H_1) &= \int  \left( \prod_{j\in L} \mathcal{N}(y_{t,j}^r; a  \bar{h}_j(s), \sigma_{r,t}^2) \right) \nonumber \\
 &\qquad \qquad  \qquad   \times p(a,s,v|H_1) \textrm{d}a \textrm{d}s \textrm{d}v \nonumber \\
 & =\int   p(y^r_{t,L}|a,s) p(a,s|H_1) \textrm{d}a \textrm{d}s  \label{eq:derivetestratio}
  \end{align}
 where $p(a,s|H_1)$ is either the marginal (or restriction to intensity and spatial position only) of the law of the birth $\mu_{\psi}$ (see \eqref{eq:state}) if proposing the initial state of the birth move, or the pdf $p(a,s|a_{0:k}, a_{0:k})$ (to be defined below) if extending the target intensity and position trajectory after having created the initial intensity and position. $p(y^r_{t,L}|a,s)$ is implicitly defined.
 
Let $p(y^r_{t,L},a,s|H_1)=p(y^r_{t,L}|a,s) p(a,s|H_1)$. Use the approximation 
 \begin{align}
 \ln p(y^r_{t,L},a,s|H_1) & \approx \ln p(y^r_{t,L}, \bar{a},\bar{s}|H_1) \nonumber \\ 
 & \quad -\frac{1}{2} [(a,s)-(\bar{a},\bar{s}) ] D [(a,s)- (\bar{a},\bar{s}) ]^T \label{eq:pseudolap}
 \end{align}
 where $-D$ is the second order derivative $\nabla^2 \ln p(y^r_{t,L},a,s|H_1)$ evaluated at $(\bar{a},\bar{s})$.  Expression \eqref{eq:pseudolap} is like the Laplace approximation except that the second order Taylor expansion is computed at $(y^f_{t,i},\Delta r, \Delta c)$ and not the true 
maximum $\arg \max_{a,s} p(y^r_{t,L},a,s|H_1)$ to save on the maximization step, which we find in the numerical examples to be still effective as a component of the birth move. (Moreover, 
it is a fair simplification for a diffused prior.)
 Thus \[ p(y^r_{t,L}|H_1)\approx p(y^r_{t,L}, \bar{a},\bar{s} |H_1) \; \frac{(2\pi)^{3/2}}{\sqrt{|D|}} \] and 
the Gaussian distribution in \eqref{eq:birthinnitstate} is 
\begin{equation} \mathcal{N}(\cdot | (\bar{a},\bar{s}),D^{-1}). \label{eq:derivgaussian}
\end{equation}

Sec.\ \ref{sec: Numerical examples} (numerical examples) assumes a linear and Gaussian model for the targets in both the synthetic and real data examples. 
The description is now completed by specifying $p(\bar{a},\bar{s}|H_1)$. 

When the birth move is constructing the initial/first state of the target, 
\begin{equation}p(a,s|H_1)= \int \mu_{\psi}(a,s,v)\textrm{d}v \label{eq:truncbirth}\end{equation}
and the expression in \eqref{eq:ModelRatio} now simplifies to 
\begin{align}
	  \rho(y_{t,L(i)}^r) =\frac{p(H_1)}{p(H_0)}\frac{p(y^r_{t,L}| \bar{a},\bar{s})}{p(y^r_{t,L}|H_0)} p(\bar{a},\bar{s}|H_1) \frac{(2\pi)^{3/2}}{\sqrt{|D|}}.  \label{eq:deriverho}
		\end{align}
Finally, we derive $p(a,s| H_1)$ in \eqref{eq:derivetestratio} when the birth move is extending the target intensity and position trajectory after having created the initial intensity position pairs $(a_{0}, s_{0}),\ldots, (a_{k-1}, s_{k-1})$ for some $k\geq 1$. For a target with a linear Gaussian model \eqref{eq:state}, the pdf of  $v_{k-1}$ (or $v_{0:k-1}$) conditioned on $s_{0:k-1}$, which is denoted $p_{\psi}(v_{k-1}|{s}_{0:k-1})$, is a Gaussian.  Thus $p(a,s| H_1)$ is 
\[\int f_{\psi}(a,s,v| a_{k-1},s_{k-1},v_{k-1}) p_{\psi}(v_{k-1}|{s}_{0:k-1}) \mathrm{d}v \mathrm{d}v_{k-1}\] and the corresponding expression for $\rho(y_{t_k,L(i)}^r)$ in \eqref{eq:birthaddmorestates} is the same as in \eqref{eq:deriverho}. 

\section{State proposal} \label{app:state}
 The state proposal of Sec.\ \ref{sec:stateproposal} alters the state values of the targets whose trajectories
have been partially exchanged. This proposal is defined for the Gaussian model in Sec.\ \ref{sec: Numerical examples}.

Let $\left\{ (k,t_{b}^{k},\mathbf{x}^{k})\right\} _{k=1}^{K}$ be
the MTT state and assume without loss of generality $U=\{1,2\}$.
The state porposal $q_{s,\theta}(U,t,\tilde{\mathbf{x}}^{1},\tilde{\mathbf{x}}^{2}|\mathbf{x}^{1:K},t_{b}^{1:K},y_{1:n})$
can be decomposed as the product of $q_{s,\theta}(U,t|\mathbf{x}^{1:K},t_{b}^{1:K})$
and $q_{s,\theta}(\tilde{\mathbf{x}}^{1},\tilde{\mathbf{x}}^{2}|\mathbf{x}^{1:K},t_{b}^{1:K},y_{1:n},U,t)$.
The first term is the probability of selecting $(U,t)$ which is not
$y_{1:n}$ dependent. Using $y_{1:n}$ and $\left\{ (k,t_{b}^{k},\mathbf{x}^{k})\right\} _{k=3}^{K}$,
generate the residual image as in \eqref{eq:imageResidual} by subtracting
the background intensity and the contribution from all targets except
$(t_{b}^{1},\mathbf{x}^{1})$ and $(t_{b}^{2},\mathbf{x}^{2}$). Let
$y_{t}^{r}=(y_{t,1}^{r},\ldots,y_{t,m}^{r})$ denote the residual
images at time $t$. The state proposal samples $(\tilde{\mathbf{x}}^{1},\tilde{\mathbf{x}}^{2})$
from the pdf $q_{s,\theta}(\tilde{\mathbf{x}}^{1},\tilde{\mathbf{x}}^{2}|\hat{\mathbf{x}}^{1:2},\tau_{b}^{1:2},y_{1:n}^{r})$
where $(\hat{\mathbf{x}}^{1},\hat{\mathbf{x}}^{2})$ is defined
in \eqref{eq:swaptraj}. Note the dependancy on targets $k>2$ is captured through
the residual image.

For brevity, we write  $({\mathbf{x}}^{1},{\mathbf{x}}^{2})$ instead of 
$(\hat{\mathbf{x}}^{1},\hat{\mathbf{x}}^{2})$. Also $\mathbf{x}^{i}=(a_{0}^{i},s_{0}^{i},v_{0}^{i},\ldots,a_{l^{i}-1}^{i},s_{l^{i}-1}^{i},v_{l^{i}-1}^{i})$
is expressed as $(\mathbf{a}^{i},\mathbf{s}^{i},\mathbf{v}^{i})$ to highlight
the intensity, spatial coordinates and velocity components.

The proposal $q_{s,\theta}$ does not alter the spatial components,
i.e. $\tilde{\mathbf{s}}^{1}=\mathbf{{s}}^{1}$ and $\tilde{\mathbf{s}}^{2}=\mathbf{{s}}^{2}$.

The velocities $(\tilde{\mathbf{v}}^{1},\tilde{\mathbf{v}}^{2})\sim p_{\psi}(\tilde{\mathbf{v}}^{1}|\mathbf{{s}}^{1})p_{\psi}(\tilde{\mathbf{v}}^{2}|\mathbf{{s}}^{2})$,
i.e. are sampled independently from $p_{\psi}$, which is the prior
pdf of the velocity conditioned on the spatial coordinates
values, which is Gaussian.

If targets 1 and 2 exist at time $t$ then their state values are
$x_{t-\tau_{b}^{1}}^{1}$ and $x_{t-\tau_{b}^{2}}^{2}$
and respectively. We assume (for pixel $i$) $y_{t,i}^{r}\sim\mathcal{N}(\cdot|0,\sigma_{r,t}^{2})$
if targets 1 and 2 both do not exist at time $t$, $y_{t,i}^{r}\sim\mathcal{N}(\cdot|a_{t-\tau_{b}^{1}}^{1}h_{i}(s_{t-\tau_{b}^{1}}^{1}),\sigma_{r,t}^{2})$
if only target 1 exists and $y_{t,i}^{r}\sim\mathcal{N}(\cdot|a_{t-\tau_{b}^{1}}^{1}h_{i}(s_{t-\tau_{b}^{1}}^{1})+a_{t-\tau_{b}^{2}}^{2}h_{i}(s_{t-\tau_{b}^{2}}^{2}),\sigma_{r,t}^{2})$
if both exists. The prior probability model (see {\eqref{eq:state}) for the
intensities are independent Gaussians, denoted $p_{\psi}(\mathbf{a}^{1})p_{\psi}(\mathbf{a}^{2})$.
Conditioned on $y_{1:n}^{r}$, $(\tau_{b}^{1},\mathbf{s}^{1})$
and $(\tau_{b}^{2},\mathbf{s}^{2})$, the posterior pdf for
the joint intensities is also a Gaussian, which is denoted by $q_{s,\theta}(\tilde{\mathbf{a}}^{1},\tilde{\mathbf{a}}^{2} | \mathbf{s}^{1:2},\tau_{b}^{1:2},y_{1:n}^{r})$.
Thus 
\begin{eqnarray*}
 q_{s,\theta}(\tilde{\mathbf{x}}^{1},\tilde{\mathbf{x}}^{2}|\mathbf{x}^{1:2},\tau_{b}^{1:2},y_{1:n}^{r}) &= &p_{\psi}(\tilde{\mathbf{v}}^{1}|\mathbf{s}^{1})
p_{\psi}(\tilde{\mathbf{v}}^{2}|\mathbf{s}^{2})\\
&& \times \;  q_{s,\theta}(\tilde{\mathbf{a}}^{1},\tilde{\mathbf{a}}^{2}|\mathbf{s}^{1:2},\tau_{b}^{1:2},y_{1:n}^{r}).
\end{eqnarray*}

\section*{Acknowledgement}
We thank  Kristina Ganzinger and Professor David Klenerman for providing the real data and the code in \cite{weimann2013quantitative} for the comparisons in Section \ref{sec:cell}, and Sinan Y{\i}ld{\i}r{\i}m for his careful reading of this paper.

\small
\bibliographystyle{IEEEtran}
\bibliography{Thesis}

\begin{thebibliography}{10}
\providecommand{\url}[1]{#1}
\csname url@samestyle\endcsname
\providecommand{\newblock}{\relax}
\providecommand{\bibinfo}[2]{#2}
\providecommand{\BIBentrySTDinterwordspacing}{\spaceskip=0pt\relax}
\providecommand{\BIBentryALTinterwordstretchfactor}{4}
\providecommand{\BIBentryALTinterwordspacing}{\spaceskip=\fontdimen2\font plus
\BIBentryALTinterwordstretchfactor\fontdimen3\font minus
  \fontdimen4\font\relax}
\providecommand{\BIBforeignlanguage}[2]{{%
\expandafter\ifx\csname l@#1\endcsname\relax
\typeout{** WARNING: IEEEtran.bst: No hyphenation pattern has been}%
\typeout{** loaded for the language `#1'. Using the pattern for}%
\typeout{** the default language instead.}%
\else
\language=\csname l@#1\endcsname
\fi
#2}}
\providecommand{\BIBdecl}{\relax}
\BIBdecl

\bibitem{Bar-Shalom_and_Fortmann_1988}
Y.~Bar-Shalom and T.~E. Fortmann, \emph{Tracking and data association}.\hskip
  1em plus 0.5em minus 0.4em\relax Boston: Academic Press, 1988.

\bibitem{weimann2013quantitative}
L.~Weimann, K.~A. Ganzinger, J.~McColl, K.~L. Irvine, S.~J. Davis, N.~J. Gay,
  C.~E. Bryant, and D.~Klenerman, ``A quantitative comparison of single-dye
  tracking analysis tools using monte carlo simulations,'' \emph{PloS one},
  vol.~8, no.~5, p. e64287, 2013.

\bibitem{mahler2007statistical}
R.~P. Mahler, \emph{Statistical multisource-multitarget information
  fusion}.\hskip 1em plus 0.5em minus 0.4em\relax Boston: Artech House, 2007,
  vol. 685.

\bibitem{streit2002multitarget}
R.~L. Streit, M.~L. Graham, and M.~J. Walsh, ``Multitarget tracking of
  distributed targets using histogram-pmht,'' \emph{Digital Signal Processing},
  vol.~12, no.~2, pp. 394--404, 2002.

\bibitem{RGM05}
M.~G. Rutten, N.~J. Gordon, and S.~Maskell, ``Recursive track-before-detect
  with target amplitude fluctuations,'' \emph{IEE Proceedings - Radar, Sonar
  and Navigation}, vol. 152, no.~5, pp. 345--352, 2005.

\bibitem{boers2004multitarget}
Y.~Boers and J.~Driessen, ``Multitarget particle filter track before detect
  application,'' \emph{IEE Proceedings-Radar, Sonar and Navigation}, vol. 151,
  no.~6, pp. 351--357, 2004.

\bibitem{punithakumar2005sequential}
K.~Punithakumar, T.~Kirubarajan, and A.~Sinha, ``A sequential monte carlo
  probability hypothesis density algorithm for multitarget
  track-before-detect,'' in \emph{Optics \& Photonics 2005}.\hskip 1em plus
  0.5em minus 0.4em\relax International Society for Optics and Photonics, 2005,
  pp. 59\,131S--59\,131S.

\bibitem{samuel2007comparison}
S.~J. Davey, M.~G. Rutten, and B.~Cheung, ``A comparison of detection
  performance for several track-before-detect algorithms,'' \emph{EURASIP
  Journal on Advances in Signal Processing}, vol. 2008, 2007.

\bibitem{vo2010joint}
B.-N. Vo, B.-T. Vo, N.-T. Pham, and D.~Suter, ``Joint detection and estimation
  of multiple objects from image observations,'' \emph{Signal Processing, IEEE
  Transactions on}, vol.~58, no.~10, pp. 5129--5141, 2010.

\bibitem{PaK15}
F.~Papi and D.~Y. Kim, ``A particle multi-target tracker for superpositional
  measurements using labeled random finite sets,'' \emph{IEEE Transactions on
  Signal Processing}, vol.~63, no.~16, pp. 4348--4358, 2015.

\bibitem{andrieu2010particle}
C.~Andrieu, A.~Doucet, and R.~Holenstein, ``Particle markov chain monte carlo
  methods,'' \emph{Journal of the Royal Statistical Society: Series B
  (Statistical Methodology)}, vol.~72, no.~3, pp. 269--342, 2010.

\bibitem{jiang_fusion2014}
L.~Jiang, S.~Singh, and S.~Yildirim, ``A new particle filtering algorithm for
  multiple target tracking with non-linear observations,'' in \emph{Information
  Fusion (FUSION), 2014 17th International Conference on}, July 2014, pp. 1--8.

\bibitem{kantas2015}
N.~Kantas, A.~Doucet, S.~S. Singh, J.~Maciejowski, and N.~Chopin, ``On particle
  methods for parameter estimation in state-space models,'' \emph{Statist.
  Sci.}, vol.~30, no.~3, pp. 328--351, 08 2015.

\bibitem{Oh_et_al_2009}
S.~Oh, S.~Russell, and S.~Sastry, ``Markov chain monte carlo data association
  for multi-target tracking,'' \emph{IEEE Trans. Automat. Control}, vol.~54,
  no.~3, pp. 481--497, Mar. 2009.

\bibitem{vu2014particle}
T.~Vu, B.-N. Vo, and R.~Evans, ``A particle marginal metropolis-hastings
  multi-target tracker,'' \emph{IEEE Trans. Signal Process}, vol.~62, no.~15,
  pp. 3953 -- 3964, 2014.

\bibitem{KoS2015}
J.~Kokkala and S.~Sarkka, ``Combining particle {MCMC} with rao-blackwellized
  monte carlo data association for parameter estimation in multiple target
  tracking,'' \emph{Digital Signal Processing}, vol.~47, pp. 84 -- 95, 2015.

\bibitem{Lan2014}
L.~Jiang, S.~Singh, and S.~Y{\i}ld{\i}r{\i}m, ``Bayesian tracking and parameter
  learning for non-linear multiple target tracking models,'' \emph{IEEE Tran.
  Signal Proc.}, vol.~63, pp. 5733--5745, 2015.

\bibitem{Duckworth:EECS-2012-68}
\BIBentryALTinterwordspacing
D.~Duckworth, ``Monte carlo methods for multiple target tracking and parameter
  estimation,'' Master's thesis, EECS Department, University of California,
  Berkeley, May 2012. [Online]. Available:
  \url{http://www.eecs.berkeley.edu/Pubs/TechRpts/2012/EECS-2012-68.html}
\BIBentrySTDinterwordspacing

\bibitem{SWG11}
S.~S. Singh, N.~Whiteley, and S.~J. Godsill, ``Approximate likelihood
  estimation of static parameters in multi-target models,'' in \emph{Bayesian
  Time Series Models}, D.~Barber, A.~T. Cemgil, and S.~Chiappa, Eds.\hskip 1em
  plus 0.5em minus 0.4em\relax Cambridge University Press, 2011, pp. 225--244.

\bibitem{yildirim2014calibrating}
S.~Y{\i}ld{\i}r{\i}m, L.~Jiang, S.~S. Singh, and T.~A. Dean, ``Calibrating the
  gaussian multi-target tracking model,'' \emph{Statistics and Computing}, pp.
  1--14, 2014.

\bibitem{jasra2005markov}
A.~Jasra, C.~Holmes, and D.~Stephens, ``Markov chain monte carlo methods and
  the label switching problem in bayesian mixture modeling,'' \emph{Statistical
  Science}, pp. 50--67, 2005.

\bibitem{whiteleydiscussion}
N.~Whiteley, ``Discussion on the paper by {A}ndrieu, {D}oucet and
  {H}olenstein.''

\bibitem{green1995reversible}
P.~J. Green, ``Reversible jump {M}arkov chain {M}onte {C}arlo computation and
  {B}ayesian model determination,'' \emph{Biometrika}, vol.~82, no.~4, pp.
  711--732, 1995.

\end{thebibliography}

\end{document}